\documentclass[notitlepage,12pt]{article}
\usepackage{fancyvrb, verbatim, listings}                                  
\usepackage{setspace}                                                      
\usepackage[margin=2.54cm]{geometry}                                       
\usepackage[UKenglish,cleanlook]{isodate}                                  
\usepackage{fancyhdr} 
\usepackage[activate={true,nocompatibility},final,tracking=true,kerning=true,spacing=true,factor=1100,stretch=10,shrink=10]{microtype}
\microtypecontext{spacing=nonfrench}                                       
\usepackage{graphicx}                                                      
\usepackage[table,dvipsnames]{xcolor}                                      
\usepackage{float}                                                         
\usepackage[justify]{ragged2e}                                             
\usepackage{booktabs}                                                      
\usepackage{multirow}                                                      
\usepackage[labelfont=bf]{caption}
\usepackage{subcaption}
\usepackage[shortlabels]{enumitem}                                         
\usepackage{tikz}
\usetikzlibrary{shapes,decorations,decorations.pathreplacing,arrows,calc,arrows.meta,fit,positioning}
\tikzset{
    auto,node distance =1 cm and 1 cm,semithick,
    state/.style ={ellipse, draw, minimum width = 0.7 cm},
    point/.style = {circle, draw, inner sep=0.04cm,fill,node contents={}},
    bidirected/.style={Latex-Latex,dashed},
    el/.style = {inner sep=2pt, align=left, sloped}
}
\usepackage{mathtools}                                                     
\usepackage{amssymb}                                                       
\usepackage{amsmath}                                                       
\usepackage{dsfont}                                                        
\usepackage{centernot}                                                     
\usepackage{siunitx} 
\usepackage{amsthm}                                                        
\renewcommand{\vec}[1]{\boldsymbol{\mathit{#1}}}                           
\newcommand{\E}[2][]{\mathbb{E}_{#1} \left[ #2 \right]}                    
\newcommand{\Egiven}[3][]{\mathbb{E}_{#1} \left[ #2 \, \middle\vert \, #3 \right]} 
\newcommand{\partialdiff}[2][]{\frac{\partial#1}{\partial#2}}              
\newcommand{\indep}{\, \raisebox{0.05em}{\rotatebox[origin=c]{90}{$\models$}} \,}
\definecolor{ao(english)}{rgb}{0.0, 0.5, 0.0}                              
\newcommand{\appendixref}[1]{\hyperref[#1]{Appendix~\ref*{#1}}}
\usepackage{natbib} 
\usepackage[backref=page]{hyperref}                                        
\hypersetup{colorlinks=true, linkcolor=blue, citecolor=blue, filecolor=magenta, urlcolor=blue}
\def\sectionautorefname~#1\null{Section~#1\null}                           
\def\subsectionautorefname~#1\null{Subsection~#1\null}                     
\def\subsubsectionautorefname~#1\null{Subsubsection~#1\null}               
\def\equationautorefname~#1\null{Equation~(#1)\null}                       
\usepackage{epigraph}                                                      

\author{Senan Hogan-Hennessy\thanks{
    This work has been supported by research grants from the Centre for the Study of Inequality and the Industrial Labour Relations School, Cornell University, conducted under UK Biobank Project \#335854.
    Thanks to Harry Patrinos for providing the collected literature review estimates (from \citealt{patrinos2025causal}).
    For helpful comments I thank
    Neil Cholli,
    Ryan Dycus,
    Benny Goldman,
    Takuma Habu,
    Jake Meyer,
    Douglas Miller,
    Lauri Kytömaa,
    Zhuan Pei,
    Evan Riehl,
    and
    David Slichter.
    I thank seminar participants at EALE Bergen NHH (2024), Cornell University (2025, 2026), and TGAC University of Wisconsin Madison (2026) for helpful discussion.
    Any comments or suggestions may be sent to me at \href{mailto:seh325@cornell.edu}{\nolinkurl{seh325@cornell.edu}}.
    } \\
    \vspace{0.1cm}
    Economics Department, Cornell University\footnote{
        Address: Uris Hall \#447, Economics Department, Cornell University NY 14853 USA.
    }
}
\title{The Direct and Indirect Effects of \\Genetics and Education}
\date{
    \vskip-0.5cm
    \today
}
\begin{document}
\clearpage
\maketitle
\thispagestyle{empty}
\begin{abstract}
    \noindent
Genes associated with educational attainment causally improve labour market income, but the economic mechanism behind this relationship is not clear.
Using quasi-random variation in genetic inheritance across siblings in the UK Biobank, I estimate the causal effect of the Education PolyGenic Index (Ed PGI) on education and income.
A one standard deviation increase in the Ed PGI raises completed education by 0.5 years and later-life income by around 5 percent (replicating the main estimates in \citealt{carvalho2024genetics}).
I then decompose this total genetic income effect into an indirect channel operating through education years and a residual direct effect, using a causal mediation framework.
Unlike structural model-based decompositions, this approach is design-based;
one remaining source of uncertainty (the causal return to an extra year of education) is handled transparently through a sensitivity analysis.
At correlational education returns of around 6 percent, 65 to 75 percent of the total genetic income effect operates through the years of education channel.
Quasi-experimental estimates from the economics literature for Britain imply higher returns to education, suggesting that the mediated share is larger and that the majority of the Ed PGI's income effect operates through completed years of education.

\vspace{0.25cm}

\vspace{0.1cm}

\end{abstract}

\newpage
\setcounter{page}{1}
\onehalfspacing
\noindent
\setlength{\epigraphwidth}{0.85\textwidth}
\vspace{-1cm}
{\singlespacing
\vspace{-1cm}
\epigraph{
    \justify
    The difference of natural talents in different men, is, in reality, much less than we are aware of [\ldots].
    The difference between the most dissimilar characters, between a philosopher and a common street porter, seems to arise not so much from nature, as from habit, custom, and education.
}{\textit{--- \citet{smith1937wealth},
    The Wealth of Nations, Volume~1 p.~16.}}}

\definecolor{childColour}{HTML}{1F77B4} 
\definecolor{motherColour}{HTML}{009E73} 
\definecolor{fatherColour}{HTML}{D55E00} 
\definecolor{parentColour}{HTML}{9370DB} 

\section{Introduction}
\label{sec:intro}
Genes associated with educational attainment causally improve labour market outcomes, but the mechanism behind this relationship remains poorly understood.
Individuals with more education-associated genes have higher incomes and more years of education, yet it is unclear how much of their income advantage reflects the additional education their genes induce and how much operates through other pathways.
Understanding the relative importance of these channels provides insight into the process of human capital formation.
If most of the income effect operates through years of completed education, then education-linked genetic variation influences labour market outcomes primarily by altering schooling decisions and educational investments, rather than through pathways outside completed years of formal education.

The Education PolyGenic Index (Ed PGI) is a weighted sum of thousands of genetic variants that together explain around 12 to 15 percent of the variation in educational attainment \citep{okbay2022polygenic}; individuals with a higher Ed PGI value have more years of completed education, and higher occupation-measured incomes --- and this relationship is causal \citep{carvalho2024genetics}.
This paper asks how much of the Ed PGI's causal effect on income operates through completed education years, and how much remains outside that channel.

This paper makes two contributions.
First, it provides an explicit quantitative decomposition of the Ed PGI's causal effect on labour-market income, separating the share operating through completed education years from a residual direct channel.
Existing research establishes that the Ed PGI affects both educational attainment and labour-market outcomes, but does not quantify the relative importance of these two pathways. Second, the paper develops a way to conduct this decomposition when education years are not randomly assigned.
The analysis combines the causal effect of the Ed PGI on education with estimates of the return to an additional year of schooling, using the correlational return as an initial benchmark and quasi-experimental estimates from the British literature to guide a sensitivity analysis.
This approach makes the central source of uncertainty explicit and shows how conclusions about the two channels change with the assumed return to education, without requiring a structural model of schooling or labour-market choices.

I use data on 24,743
 individuals from the UK Biobank to estimate the causal effect of the Ed PGI on education and labour market income, following the design of \cite{carvalho2024genetics}.
The Ed PGI varies quasi-randomly across individuals: when a parent carries two different copies of a genetic variant, which copy the child inherits is determined by a random biological draw.
I control for imputed parental Ed PGI values, constructed from the genetic data of siblings also enrolled in the UK Biobank \citep{young2022mendelian}, to isolate this random inheritance component.
A one standard deviation increase in the Ed PGI increases years of education completed by 0.5 years and raises later life income by around 5 percent, confirming the main findings in \cite{carvalho2024genetics}.

The first part of the analysis establishes the two causal effects required for the decomposition: the effect of the Ed PGI on education and its total effect on labour-market income.
The research design and these total-effect estimates build on \citet{carvalho2024genetics}, but they provide the necessary empirical foundation for the paper's causal mediation analysis.
After establishing these inputs, the second part quantifies how much of the income effect operates through completed education years.

I apply a causal mediation framework to quantify how much of the Ed PGI's causal effect on income operates through completed education years.
The decomposition combines the Ed PGI's average causal effect on education years with the return to an additional year of education.
The latter is not identified by the genetic design, so I first use the correlational return as a benchmark, and then evaluate the decomposition over the distribution of quasi-experimental returns estimated in the British economics literature \citep{patrinos2025causal}.
This approach makes the remaining source of uncertainty transparent without requiring a structural model of education or labour-market choices.

My main finding is that the Ed PGI raises labour market income primarily through education years.
At correlational returns to education of around 6 percent, roughly 65 to 75 percent of the total genetic income effect is mediated through the education years channel.
The sensitivity analysis then evaluates this decomposition over the distribution of returns to education estimated in the British economics literature.
Across this range, the mediated share rises further as the return to education increases, implying that the majority of the Ed PGI's labour market effect operates through educational attainment rather than residual non-years-of-education pathways.

Modern genetic data provide a new way to study a central question in the economics of human capital formation: how initial endowments interact with environments and investments to shape education and later-life outcomes \citep{cunha2007technology,beauchamp2011moecular,biroli2025economics}.
This paper contributes to a growing literature applying these data to socioeconomic outcomes.
\cite{papageorge2020genes} and \cite{bryson2025gender} establish that the Ed PGI predicts labour-market outcomes, although \citet{schork2022indirect} shows that these associations may partly reflect parental genetic channels rather than causal effects of an individual's own genes.
Using the sibling-imputation procedure developed by \citet{young2022mendelian}, \citet{carvalho2024genetics} addresses this concern by controlling for imputed parental PGIs and estimating causal effects; \citet{muslimova2025environment,sanz2025sibling} exploit related within-family variation to study complementary questions.
A parallel literature uses structural models of education and labour-market choices to examine how the Ed PGI shapes socioeconomic outcomes \citep{barth2022genetic,houmark2024nurture,bolt2025genetic}.
Together, these studies establish that the Ed PGI predicts and causally affects socioeconomic outcomes, and structural models provide interpretations of the choices underlying these relationships.
They do not, however, provide a design-based quantitative decomposition of the income effect into the component operating through completed education years and the component remaining outside that channel.

Causal mediation decomposes an average treatment effect into an observed mechanism and a remaining direct effect, where credible identification requires quasi-random variation in the mediating channel \citep{imai2010identification} --- see \cite{huber2019review} for an overview of the causal mediation literature.\footnote{
    Analyses of mechanisms in applied economics typically rely on suggestive evidence instead, which does not identify or quantify the role of a mechanism.
    See \cite{blackwell2024assumption,green2010enough}.
}
When the mediating channel is not randomly assigned (as education years plainly are not), conventional mediation estimates are not causal and suffer from selection bias \citep{mediation-natural-experiment}.
I instead combine the identified causal effect of the Ed PGI on education with estimates of the return to an additional year of schooling.
The correlational return provides an initial benchmark, while a sensitivity analysis shows how the decomposition changes with different values for causal returns to education.
I discipline this analysis using the distribution of quasi-experimental estimates from the economics literature for Britain.
The resulting design-based decomposition quantifies the contribution of education years without requiring a structural model of schooling or labour-market choices.

This paper proceeds as follows.
\autoref{sec:data} introduces the UK Biobank data and the construction of the Ed PGI.
\autoref{sec:genetics} and \autoref{sec:genetic-effects} present the framework and empirical results, respectively, of the two inputs to the causal mediation decomposition: the Ed PGI's effect on education and its total effect on labour-market income.
\autoref{sec:direct} develops the framework for separating the education-years and residual channels, and \autoref{sec:direct-effects} presents the empirical results, and a brief discussion.
\autoref{sec:conclusion} concludes.

\section{Data and Genetic Context}
\label{sec:data}
This paper draws on data from the UK Biobank, a large-scale cohort study of the British population with both rich socioeconomic measures and detailed genetic data.
I construct the Education PolyGenic Index (Ed PGI) summarising each participant's genetic predisposition toward educational attainment, and impute parental PGI values using genetic data from siblings.
Together, these data enable a within-family research design that separates the role of inherited genetic factors from the shared family environment in shaping socioeconomic outcomes.

\subsection{UK Biobank (UKB)}
\label{sec:data-UKB}
The UK Biobank (UKB) is a large-scale prospective cohort study comprising approximately half a million members of the British public, recruited between 2006 and 2010.
Letters were sent to a random sample of addresses in the National Health Service registry, and the final UKB data represent voluntary respondents.
Participants, aged 40-69 years at recruitment, underwent comprehensive baseline assessments, were asked many questions for socioeconomic information either by interview, automated survey answers, or follow-up sessions.
Key socioeconomic variables include educational attainment (measured in education qualifications), household income brackets, employment status, and industry of employment.
See \cite{sudlow2015uk} for a full description of the UKB organisation and data.

\begin{table}[!htp]
    \vskip-0.75cm
    \singlespacing
    \centering
    \small
    \caption{Descriptive Statistics, UKB.}
    \vskip-0.25cm
    \begin{tabular}{l c c c c}
        \\[-1.8ex]\hline \hline \\[-1.8ex] 
            & \multicolumn{2}{c}{Analysis sample}
            & \multicolumn{2}{c}{Entire sample} \\
        \cmidrule(lr){2-3} \cmidrule(lr){4-5}
        & Mean & SD & Mean & SD \\
        \\[-1.8ex]\hline \\[-1.8ex]
 \\[-1.8ex] \textit{Demographics:} &   &   &   &   \\ 
  Male & 0.44 & 0.50 & 0.46 & 0.50 \\ 
  Age & 55.04 & 7.05 & 56.55 & 8.09 \\ 
  Lives in city & 0.86 & 0.35 & 0.85 & 0.35 \\ 
  Ethnicity $=$ European & 1.00 & 0.00 & 0.82 & 0.38 \\ 
  Any siblings also in UKB? & 1.00 & 0.00 & 0.08 & 0.28 \\ 
  Count of siblings in UKB & 1.09 & 0.32 & 0.09 & 0.32 \\ 
  \\[-1.8ex]\hline \\[-1.8ex] \textit{Genetic Measures:} &   &   &   &   \\ 
  Ed PGI & 0.04 & 1.00 & 0.00 & 1.00 \\ 
  Ed PGI, imputed parental mean & -0.02 & 0.75 &   &   \\ 
  Other PGI: ADHD & -0.02 & 1.00 & -0.00 & 1.00 \\ 
  Other PGI: Asthma & -0.01 & 1.00 & 0.00 & 1.00 \\ 
  Other PGI: Bipolar & -0.01 & 1.00 & -0.00 & 1.00 \\ 
  Other PGI: BMI & -0.02 & 0.99 & 0.00 & 1.00 \\ 
  Other PGI: Diabetes (type 2) & -0.02 & 1.00 & 0.00 & 1.00 \\ 
  Other PGI: Height & 0.00 & 1.01 & -0.00 & 1.00 \\ 
  Other PGI: Schizophrenia & -0.04 & 1.00 & 0.00 & 1.00 \\ 
  \\[-1.8ex]\hline \\[-1.8ex] \textit{Education:} &   &   &   &   \\ 
  Education years & 13.84 & 3.36 & 13.72 & 3.48 \\ 
  Age left education & 16.67 & 2.47 & 16.51 & 3.04 \\ 
  Qualification, University degree & 0.33 & 0.47 & 0.33 & 0.47 \\ 
  Qualification, A-Levels & 0.46 & 0.50 & 0.45 & 0.50 \\ 
  Qualification, GCSEs & 0.76 & 0.43 & 0.71 & 0.45 \\ 
  Qualification, Professional degree & 0.32 & 0.47 & 0.29 & 0.45 \\ 
  Qualification, Vocational degree & 0.23 & 0.42 & 0.19 & 0.39 \\ 
  Qualification, No official qualifications & 0.13 & 0.34 & 0.17 & 0.38 \\ 
  \\[-1.8ex]\hline \\[-1.8ex] \textit{Labour Market Outcomes:} &   &   &   &   \\ 
  Occupation hourly wage, \pounds & 21.36 & 9.15 & 22.02 & 9.49 \\ 
  Occupation annual income, thousands \pounds & 30.27 & 19.02 & 31.95 & 20.04 \\ 
  Average hours worked, per week & 34.73 & 12.58 & 35.31 & 12.70 \\ 
  Household income, $< \pounds 18k$ & 0.13 & 0.33 & 0.20 & 0.40 \\ 
  Household income, $\pounds 18-31k$ & 0.23 & 0.42 & 0.22 & 0.41 \\ 
  Household income, $\pounds 31-52k$ & 0.28 & 0.45 & 0.22 & 0.42 \\ 
  Household income, $\pounds 52-100k$ & 0.22 & 0.41 & 0.17 & 0.38 \\ 
  Household income, $\pounds 100k <$ & 0.05 & 0.22 & 0.05 & 0.21 \\ 
  Household income, thousands $\pounds$ (midpoint imputed) & 67.13 & 40.07 & 60.95 & 41.46 \\ 
  
        \\[-1.8ex]\hline \\[-1.8ex]
        Observations
            & 24,743
 &
            & 496,004
 & \\
        \\[-1.8ex]\hline \\[-1.8ex]
    \end{tabular}
    \vspace{-0.25cm}
    \label{tab:ukb-summary}
    \justify
    \footnotesize
    \textbf{Note:}
    This table shows the Mean and Standard Deviation (SD) of UKB variables, comparing the analysis subsample and the entire UKB sample.
    The analysis sample refers to individuals with at least one family member also included in the UKB (sibling or parent), have non-missing education and labour market occupation codes, and are of European ancestry.
    All genetic variables are coded as missing among those who report ancestry that is not European.
\end{table}

The UKB collected detailed information on educational qualifications, including completion of primary school, GCSEs (typically completed at age 16), A-Levels (completed at age 18), and completion of an higher education degree.
I convert these qualifications into education years based on each participant's highest reported qualification, following the International Standard Classification of Education.\footnote{
    British vocational degrees are inconsistently coded by the International Standard Classification of Education.
    I follow \cite{okbay2022polygenic}, and assign education years as the age they reported leaving full-time education minus five, among those with a vocational degree as their highest qualification.
}
While the UKB collected household income data, this information was recorded only in categorical bins; I assign midpoint imputed values and update to 2024 GBP for a coarse measure of household income.
To accompany these measures with a finer measure of income, I also calculate occupation income.
I construct an individual-level income measure using participants' 4-digit Standard Occupational Classification (SOC) codes.
Specifically, I impute occupation-based wages by estimating hourly earnings for each SOC category using data from the British Annual Survey of Hours and Earnings, conditional on age, gender, and year.\footnote{
    This approach uses replication data files provided by \cite{kweon2025associations}, also used by \cite{carvalho2024genetics}; see \appendixref{appendix:pgi-impute}.
}
This approach provides a more granular and continuous measure of economic status than the categorical household income variable.
I create the annual occupation income by multiplying this by the number of hours they reported working, times 52 weeks in the year.
This gives 496,004
individuals in the UKB, with non-missing data in relevant variables.

\begin{figure}[!h]
    \centering
    \singlespacing
    \caption{Distribution of UKB Education Measures.}
    \begin{subfigure}[b]{0.495\textwidth}
        \centering
        \caption{Education Years.}
        \includegraphics[width=\textwidth]{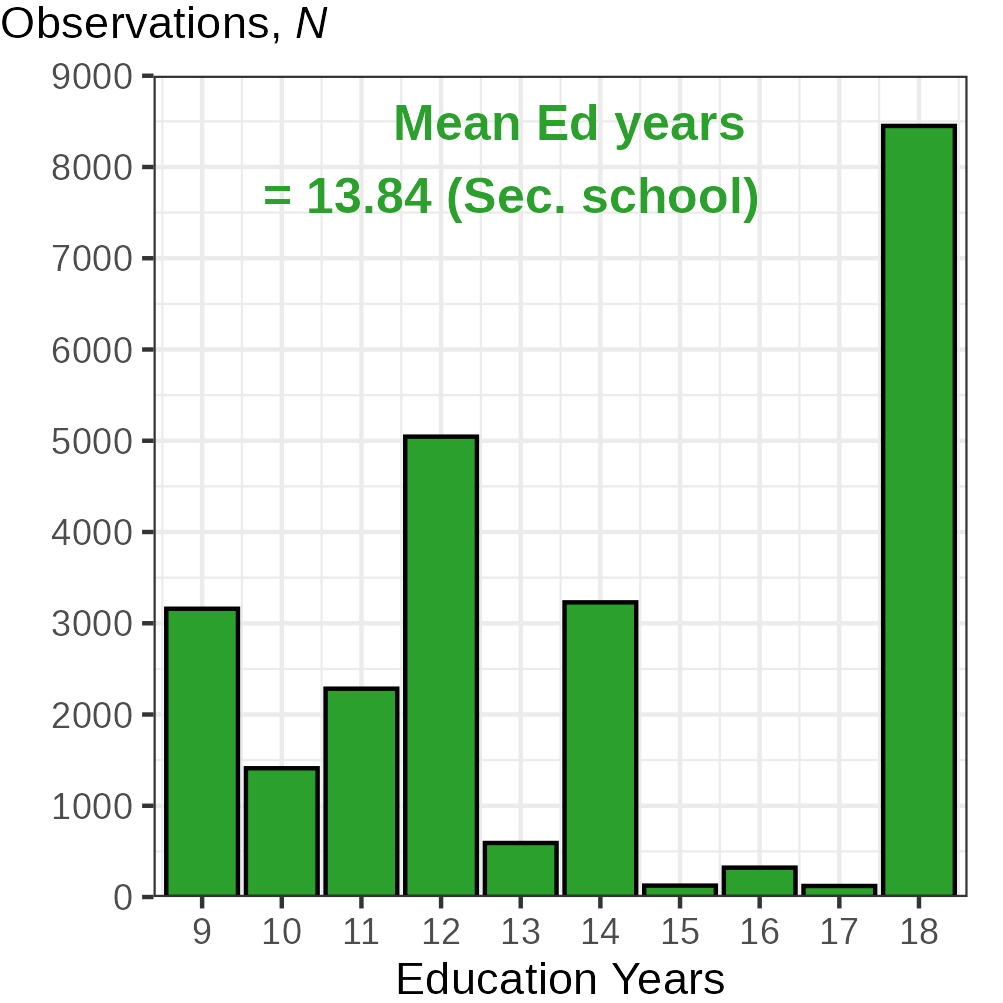}
        \label{fig:edyears-hist}
    \end{subfigure}
    \begin{subfigure}[b]{0.495\textwidth}
        \centering
        \caption{Education Years $+$ Income.}
        \includegraphics[width=\textwidth]{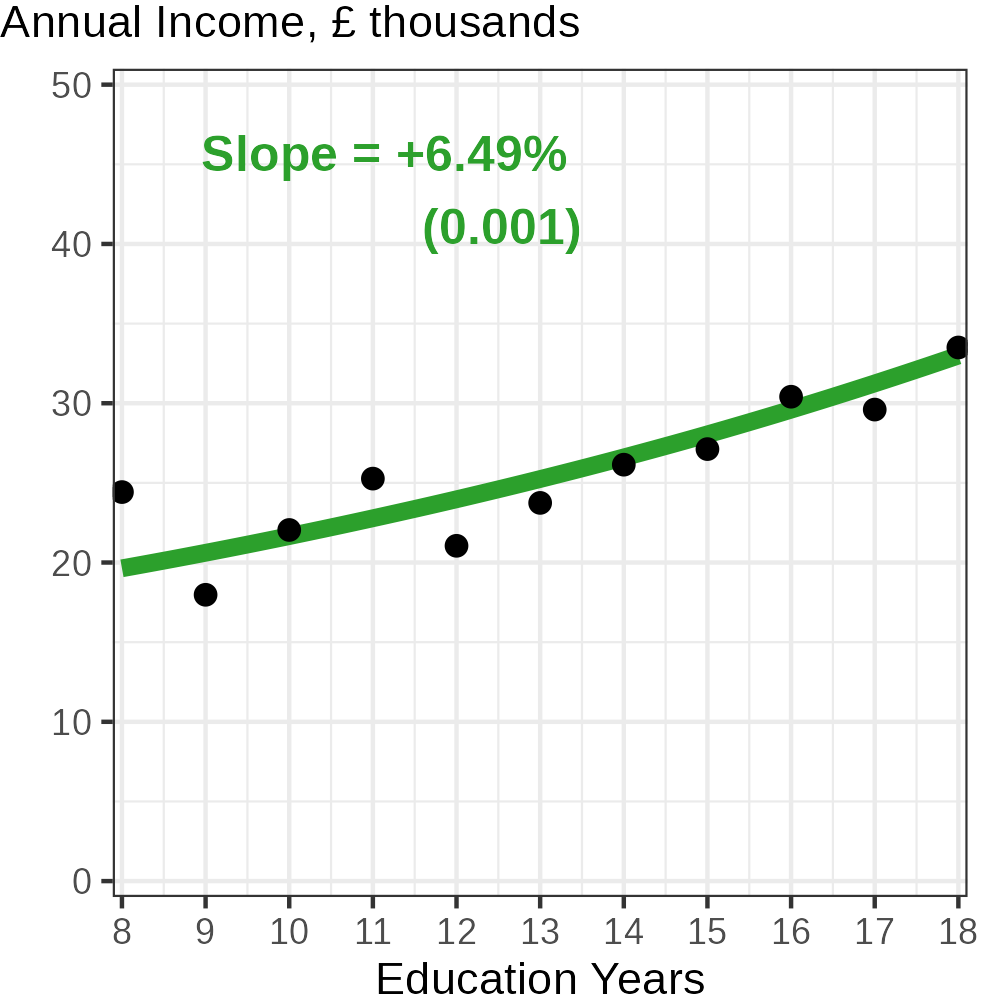}
        \label{fig:edyears-earnings}
    \end{subfigure}
    \label{fig:ukb-edyears-hist}
    \vspace{-1cm}
    \justify
    \footnotesize
    \textbf{Note}:
    The left plot shows the observation count in the UKB analysis sample with each number of reported education years, coded from education qualifications.
    The right plot shows the mean occupation imputed annual income for each level of education years, with a line of best fit calculated from a linear regression (with annual income measured in a log specification); the slope refers to the estimated coefficient from this linear regression.
\end{figure}

For the purposes of this study, the primary analysis sample is restricted to those with at least one family member, sibling or parent,\footnote{
    The UKB does not personally identify participants, nor their family members, so I follow the genetic literature by identifying family clusters by measures of genetic similarity.
    \appendixref{appendix:identifying-families} further describes the family definitions in the UKB. 
}
also enrolled in the UKB, and non-missing data in occupation code plus education qualifications.
This gives 24,743
individuals in the UKB. 
The presence of multiple family members within the dataset provides a unique opportunity to control for shared genetic factors that might otherwise bias estimates in standard cross-sectional analyses.

The analysis sample is also restricted to individuals who report to have primarily European ancestry.
The genetic measures used in this paper have generally been constructed from primarily white, European descended populations, and have little predictive power out of sample, so are only relevant in this subsample.
It is an unfortunate fact that systematic genetic measures, such as those considered in this paper, have only been constructed from Biobank samples where the vast majority are white and/or European descended; see \cite{martin2017human} for a further discussion.
A limitation of this study is that its results only apply to this subset of the UKB population.

The only notable differences between the analysis subsample and the entire UKB participant sample are that the analysis sample has more siblings, and is entirely European descended; no other variables are significantly different for the analysis subsample.
\autoref{tab:ukb-summary} summarises the UKB data, for the analysis sample and among the entire UKB sample.
\autoref{fig:ukb-edyears-hist} summarises the correlation between education years and occupation coded income.

\subsection{Education PolyGenic Index (Ed PGI)}
\label{sec:ed-pgi}
UKB participants also provided saliva samples that were subsequently genotyped, producing standardised genetic data files for statistical analysis of the relationship between genetics and outcomes.

While genetic data provide rich information about individual variation, the relationship between specific genes and social outcomes like education is extraordinarily complex.
The biological pathways from DNA to observable traits remain largely opaque --- the exact relationship between a single gene, its resulting protein (or its moderating role), and the resulting outcomes are only known for a small number of genes in the human genome.
For example, a single gene (and its recessive variant) is known to lead to a protein which does not break down the by-products of alcohol digestion; individuals with this genetic variant generally do not drink alcohol at all because of the biological limit on digestion \citep{millwood2023alcohol}.
Complex social behaviours, on the other hand, have no single gene with a known biological pathway, so cannot be summarised with one single gene, nor single biological pathway.
Instead, complex social behaviours are a result of thousands of genes, and their interactions with the environment --- they are polygenic.

Rather than attempting to trace all of these intricate biological mechanisms, researchers have developed PolyGenic Indices (PGIs) as dimensional reduction tools that summarise the statistical relationship between thousands of genetic variants and outcomes of interest.
This statistical aggregation is particularly valuable because individual genetic variants typically have tiny associations with social outcomes like educational attainment --- often explaining less than 0.01\% of the variation.
However, by combining information across thousands of genetic variants, each contributing a small predictive association, PGIs can capture meaningful predictive power of genes for some social outcomes.
These indices are fundamentally correlational measures: they capture statistical associations between genetic variants and outcomes observed in large samples, not well understood biological pathways.

The Education PGI (Ed PGI) is the PolyGenic Index built to predict years of completed education, calculated from weights derived from a Genome-Wide Association Study (GWAS), which analysed the correlation between genetic variants and education years.
In a GWAS, researchers test the correlation between each Single Nucleotide Polymorphism (SNP) --- a one-letter change in the DNA sequence at a specific location --- and the outcome of interest \citep{uffelmann2021genome}.
I use the weights calculated in the GWAS conducted by \cite{okbay2022polygenic}, which used a sample of over 3 million individuals of European ancestry, primarily from 23\&Me data.

The Ed PGI is defined as the weighted sum of whether individual $i$ has the relevant SNP at each DNA location $j$,
\[ Z_i = \sum_{j = 1}^{J} w_j x_{i,j} \]
where $w_j$ are the Ed PGI weights (provided in the supplementary data to \citealt{okbay2022polygenic}), and $x_{i,j}$ is an allele count --- whether individual $i$ has the relevant SNP at DNA location $j$.\footnote{
    Every person has two copies of each chromosome, so that $x_{i,j}$ takes values 0, 1, 2 corresponding to how many copies of the SNP individual $i$ has inherited.
    The genetics literature has generally accepted that complex social behaviours are explained by additive effects (see \citealt{hill2008data}), and \cite{okbay2022polygenic} found little evidence for dominance in the Ed PGI SNPs, so that the Ed PGI makes sense as a weighted sum.
}
The \cite{okbay2022polygenic} GWAS found that around 4,000 individual SNPs were significant in predicting education years, after passing a battery of biostatistical tests and finding no evidence for significant interactions.
The resulting Ed PGI is standardised to have mean zero and standard deviation one within the entire sample, making it interpretable as the number of standard deviations from the population mean.
The Ed PGI captures approximately 12--15\% of the variation in educational attainment, making it one of the most predictive PGIs for any complex behavioural trait to date.

\begin{figure}[!h]
    \centering
    \singlespacing
    \caption{Distribution of Education PolyGenic Index (Ed PGI).}
    \begin{subfigure}[b]{0.495\textwidth}
        \centering
        \caption{Ed PGI.}
        \includegraphics[width=\textwidth]{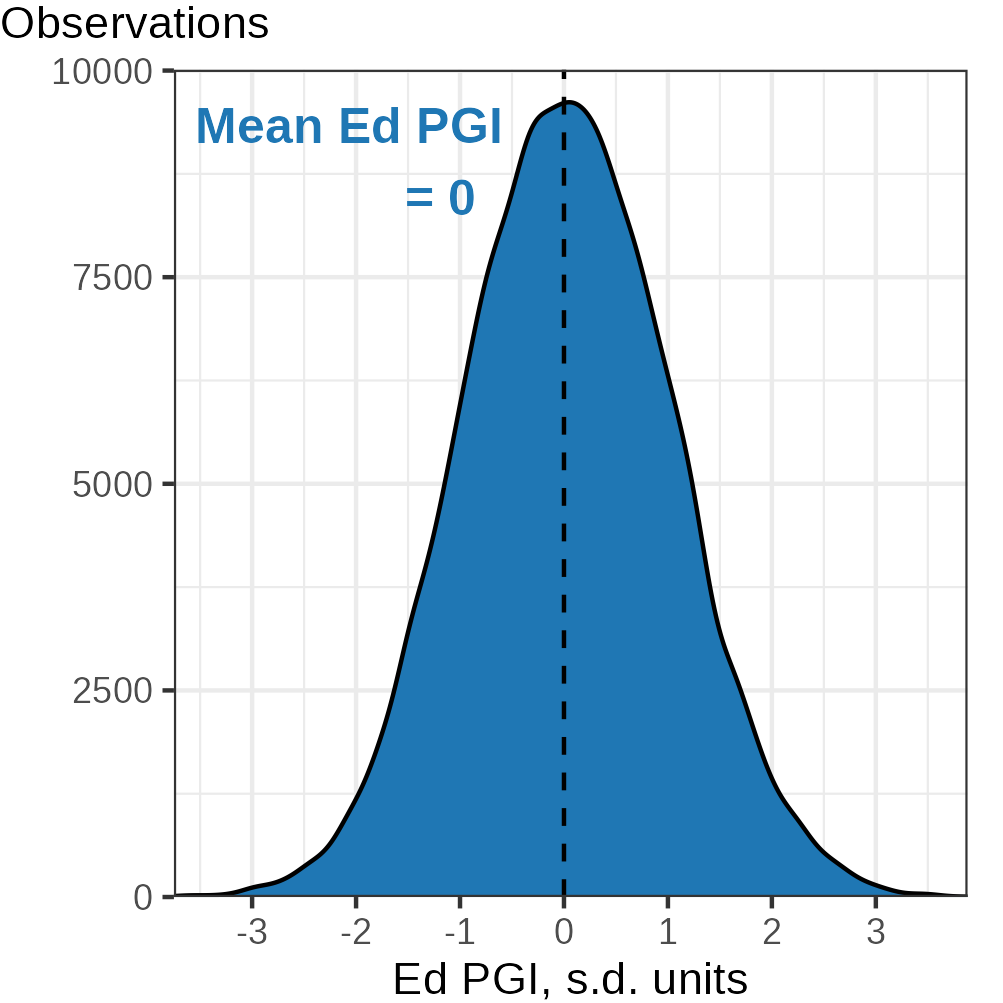}
        \label{fig:edpgi-hist}
    \end{subfigure}
    \begin{subfigure}[b]{0.495\textwidth}
        \centering
        \caption{Ed PGI $+$ Education Years, Bin Scatter.}
        \includegraphics[width=\textwidth]{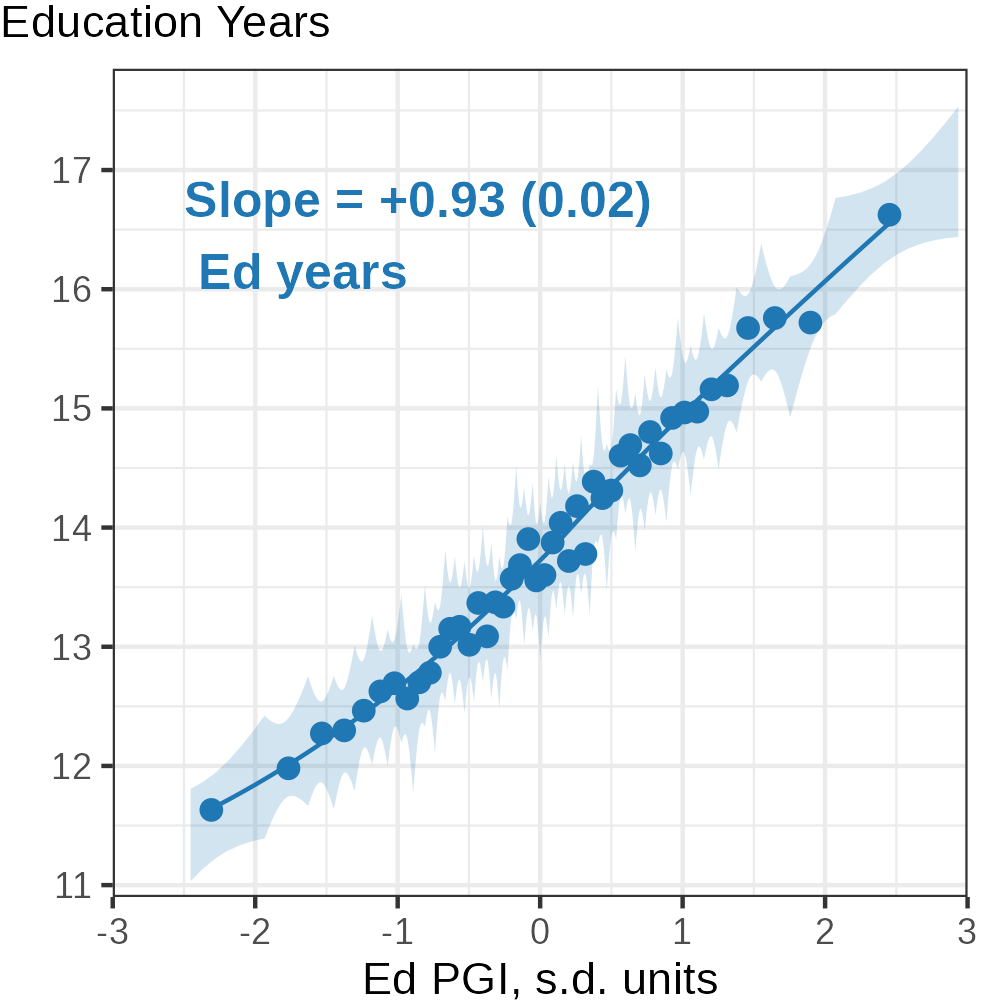}
        \label{fig:edpgi-edyears}
    \end{subfigure}
    \label{fig:ukb-hist}
    \vspace{-1cm}
    \justify
    \footnotesize
    \textbf{Note}:
    The left plot shows a kernel density estimate for the distribution of the Ed PGI among the UKB analysis sample.
    Ed PGI is normalised to have mean 0 and standard deviation 1 among the entire UKB sample; it is roughly normally distributed regardless of this normalisation.
    The right plot shows a bin scatter, visualising a (semi-parametric) correlation between the Ed PGI and education years.
    The slope refers to the estimated coefficient from a linear regression between the Ed PGI and education years.
\end{figure}

If education had no relationship with SNP variants, then we would expect the correlation between the Ed PGI and education years to be indistinguishable from zero; it is, however, highly correlated with education years in the UKB analysis sample.
A one standard deviation increase in the Ed PGI is associated with 0.93 extra education years (standard error 0.02), with an $R^2$ value of 0.10 for the raw regression of education years on Ed PGI.
It is roughly normally distributed, with standard deviation normalised to 1.
This distribution is not surprising, as the Ed PGI is the weighted sum of a few thousand binary variables, so we should expect convergence towards a normal distribution among the population.
The main takeaway is that the correlation between a summary measure of genetic values is strong, and has meaningful predictive power for education years.

The Ed PGI is not the only PGI that can be constructed.
Following the same methodology, I construct additional PGIs for seven health-related traits that may confound the relationship between genetics and socioeconomic outcomes: attention deficit hyperactive disorder (ADHD), asthma, bipolar disorder, Body Mass Index (BMI), height, schizophrenia, and type 2 diabetes.
Each PGI is calculated using the same approach as the Ed PGI --- taking weights from a published GWAS, conducted on independent populations, and computing weighted sums of the relevant genetic variants in the UKB data.
These other PGIs serve as control variables in my analysis for two reasons.
First, genetic variants often exhibit pleiotropy, meaning a single variant can influence multiple traits simultaneously.
For instance, genetic variants associated with education might also affect other factors (e.g., mental health conditions), which could independently influence labour market outcomes.
Second, systematic differences in gene frequencies across ancestry groups can create spurious correlations between any PGI and outcomes if health conditions vary across these groups for unrelated reasons.

\subsection{Imputing Parental PGIs from Siblings}
\label{sec:pgi-impute}
It would be ideal if the UKB contained genetic information for the parents of every UKB participant.
While there are 6,239 participants with a parent also included in the UKB, there are only 72 with both parents.
To overcome this missing data problem, I also use siblings' genetic data to impute the mean parental PGIs from the sample of 24,743
 participants with at least one family member (parent or sibling) also in the UKB --- the analysis sample described above.
The majority of this analysis sample have genetic data on one sibling member, so I refer to this as sibling imputation.

\begin{figure}[!h]
    \centering
    \singlespacing
    \caption{Imputing One SNP for Parents from Sibling Data.}
    \label{fig:pgi-imputing}
    \input{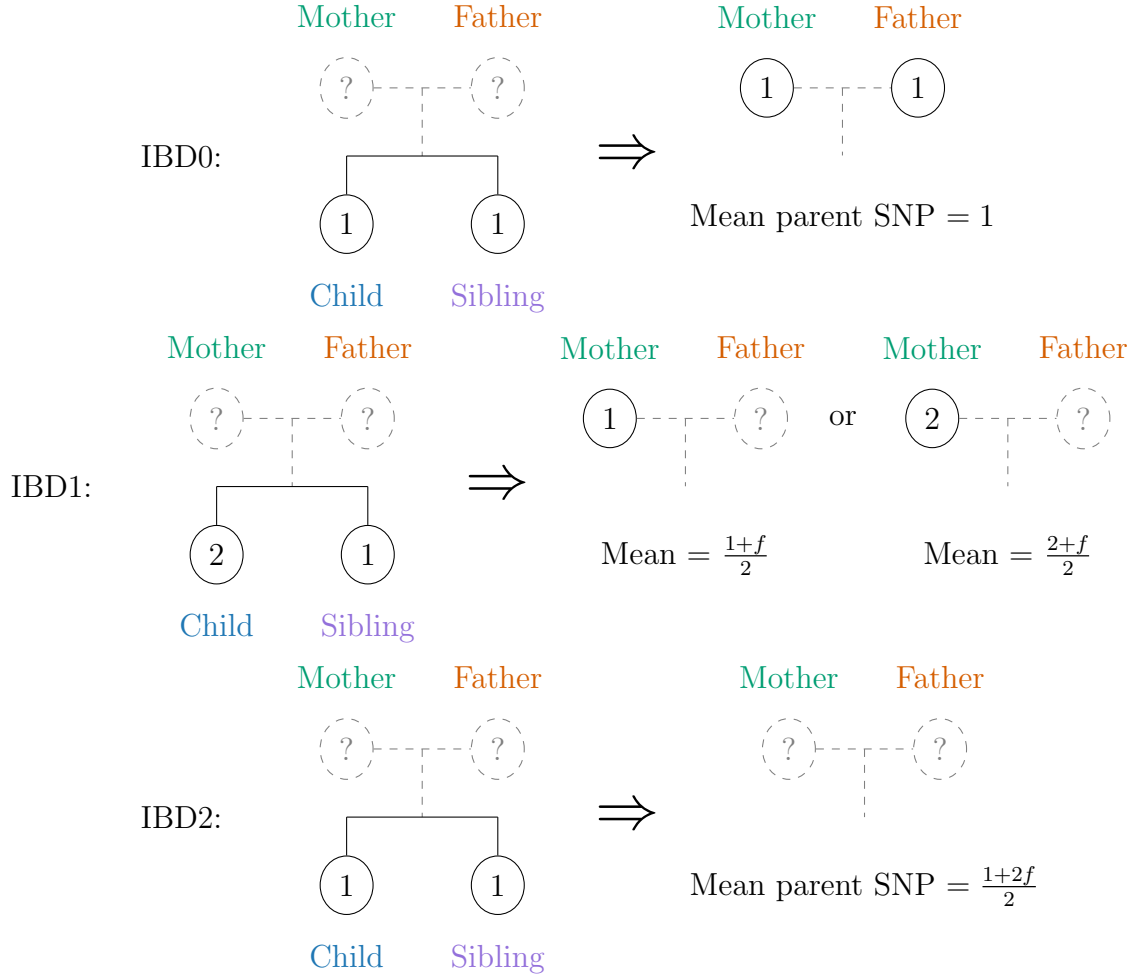}
    \justify
    \footnotesize
    \textbf{Note}:
    This figure illustrates the logic of imputing the sum of parental genotypes from sibling data at a single SNP, following \cite{young2022mendelian}.
    At each SNP, individuals have a genotype of 0, 1, or 2, representing the number of copies of a reference allele inherited from their parents.
    This figure illustrates the imputation of the parental genotype sum under the three possible identity-by-descent (IBD) states, using informative sibling-genotype configurations for each state.
    In IBD0, the siblings share no alleles by descent, so all four parental alleles are directly observed and the parental sum is known exactly as 2.
    In IBD1, the siblings share one allele by descent, leaving one parental allele unobserved; the imputed parental sum is $1 + f$ or $2+f$, depending on whether the shared allele is the reference allele, where $f$ denotes the population frequency of the reference allele.
    In IBD2, the siblings share both alleles by descent, so only two parental alleles are observed and the imputed parental sum is $1 + 2f$.
    In all cases, unobserved parental alleles are replaced by the population allele frequency $f$.
\end{figure}

I use the \cite{young2022mendelian} procedure to impute mean parental PGI values using genetic data on participants' siblings.
The key insight is that since siblings both inherit their DNA from the same two parents, observing two siblings' genotypes allows us to partially reconstruct the parental genotypes they were drawn from.
At each SNP, the imputation proceeds in two steps.
First, the procedure classifies each sibling pair as being in one of three identity-by-descent (IBD) states using a hidden Markov model that exploits information across all SNPs: IBD0, in which the siblings share no alleles inherited from the same parent; IBD1, in which they share one such allele; and IBD2, in which they share both.
This IBD classification has been shown to be correct 99.65\% of the time in validation samples \citep{young2022mendelian}.
Second, given the inferred IBD state, the procedure imputes the sum of the two parental genotypes at that SNP --- which is the quantity needed to construct a parental PGI --- by summing the directly observed parental alleles and replacing any unobserved alleles with the population allele frequency $f \in [0,1]$.

\autoref{fig:pgi-imputing} illustrates this logic for the case where both the child and their sibling have observed SNP value of 1 (one copy of the reference allele).
In IBD0, the siblings share no alleles, so all four parental alleles are directly observed across the two children, and the parental sum is known exactly as 2.
In IBD1, one allele is shared between the siblings, leaving one parental allele unobserved; depending on whether the shared allele is the reference allele or not, the imputed parental sum is either $1 + f$ or $2 + f$, depending on value of the shared  inherited allele.
In IBD2, both siblings inherited the same two alleles, so only two of the four parental alleles are observed (summing to 1), and the two unobserved alleles are each imputed as $f$, giving sum of parents $1 + 2f$.
In all cases, the imputed value is the conditional expectation of the parental sum given the observed sibling genotypes and IBD state, where unobserved alleles are replaced by the population frequency.
Averaging across the three IBD states --- which occur with probabilities 1/4, 1/2, and 1/4 respectively --- siblings jointly reveal on average three of the four parental alleles, meaning approximately 75\% of parental genetic information is directly observed rather than imputed.
See \cite{young2022mendelian} for further reasoning on the imputation performance relative to observed parents in the UKB.

\subsection{Ethical Considerations}
\label{sec:ethics}
The genetic measures in this study should not be misinterpreted as evidence of genetic determinism.
Having a higher Ed PGI does not destine someone to higher education or income; it represents a statistical tendency, not fate.
These tendencies are fundamentally shaped by social policies, institutional structures, and cultural contexts that vary across time and place.
Just as eyesight correcting glasses  can eliminate the practical consequences of genetic variation in eyesight, education and economic policies can modify or eliminate the consequences of genetic variation in traits relevant to socioeconomic outcomes.

Furthermore, this research provides no support for ideas about ``natural'' social hierarchies or genetic justifications for inequality.
In addition, genetic associations identified within one ancestry group cannot be meaningfully compared across groups due to different genetic architectures, environmental contexts, and historical experiences.
Any claim using the measures considered in this paper for inference across racial groups would be both invalid and ethically deficient.
The appropriate interpretation of these results is that they highlight how both social and genetic factors influence socioeconomic outcomes, which should inspire greater solidarity and support for policies that promote equality of opportunity rather than acceptance of inequality as ``natural'' or inevitable.

\section{Random Variation in Genetic Inheritance}
\label{sec:genetics}
Genetic variants are not randomly assigned, which means that simple correlations between the Ed PGI and observed outcomes cannot straightforwardly be interpreted as causal.
Genetic variation arises from inheritance: at conception, individuals inherit 23 chromosomes from each parent, obtaining half of their genetic variants from their mother and half from their father.
Those who inherit a higher Ed PGI therefore typically also inherit familial environments shaped by parents who themselves possess a higher Ed PGI.
Consequently, correlations between an individual's Ed PGI and outcomes may reflect effects operating through the family environment.
However, inheritance also contains a random component: conditional on parental genetics, Mendelian inheritance generates random variation in the genetic variants a child receives.
I use this variation to estimate the causal effects of the Ed PGI.

This research design establishes the two causal effects used in the later mediation analysis: the Ed PGI's effect on education and its total effect on labour-market income.
The design and resulting estimates build on existing work rather than constituting the paper's main new contribution.
Nevertheless, establishing them correctly is essential because they provide the empirical inputs to the design-based decomposition that follows.
The causal mediation analysis uses these causal estimates to separate the total income effect into the component operating through completed education years and the component remaining outside that channel.

\subsection{Mendelian Independent Inheritance}
I use variation away from parental values to estimate causal effects of having a higher versus lower Ed PGI.
This approach leverages the fundamental randomness in genetic inheritance, first described by Mendel's laws \citep{mendel1865versuche}.

Each parent has 0, 1, or 2 copies of each genetic variant --- corresponding to presence on each of their chromosomes.
When a father and mother's sex cells are formed, exactly one of the two copies is sorted into each reproductive cell, and is then inherited by their resulting child.
A parent with one copy therefore passes it on half the time, whereas a parent with 0 or 2 copies transmits their genetic variant with certainty.
At conception, the child receives one draw from the mother and one from the father, so the child's copy-count is simply the sum of two independent draws.

Write $x_{i,j}^\text{\textcolor{childColour}{Child}}$ for the number of copies of variant $j$ that individual $i$ has (0, 1, or 2), and 
$x_{m(i),j}^\text{\textcolor{motherColour}{Mother}}, x_{f(i),j}^\text{\textcolor{fatherColour}{Father}}$ for their respective parents $m(i), f(i)$.
If the mother has two copies of variant $j$ then they bequeath one copy to their child with certainty; if the father has one, then they bequeath one copy with probability $\frac 12$; if no copies, then they bequeath none with certainty.
The child then receives their other copy from their father, by the same process, giving the following representation in expectation,
\[ \Egiven{x_{i,j}^\text{\textcolor{childColour}{Child}}}{
    x_{m(i),j}^\text{\textcolor{motherColour}{Mother}}, x_{f(i),j}^\text{\textcolor{fatherColour}{Father}}}
    = \frac 12 \Big( x_{m(i),j}^\text{\textcolor{motherColour}{Mother}}
        + x_{f(i),j}^\text{\textcolor{fatherColour}{Father}} \Big). \]

After conception, the child has inherited genetic variants from both parents --- one of the four possible discrete realisations, not the expected value.
This result came about purely from random chance, conditional on at least one parent having exactly one copy of the SNP, $x_{m(i),j}^\text{\textcolor{motherColour}{Mother}} = 1$ and/or $x_{f(i),j}^\text{\textcolor{fatherColour}{Father}} = 1$.\footnote{
    Note both parents having 0 or 2 copies of variant $j$ leaves $x_{i,j}^\text{\textcolor{childColour}{Child}} = \frac 12 \big( x_{m(i),j}^\text{\textcolor{motherColour}{Mother}}
        + x_{f(i),j}^\text{\textcolor{fatherColour}{Father}} \big)$, so that individual $i$ inherited no random variation at variant $j$.
}

\subsection{Mendelian Independent Inheritance in the Ed PGI}
\label{sec:mendel-pgi}
Moving from a single variant to the PGI simply means summing across the same random-segregation for all the variants.
The Ed PGI is constructed by summing the 0, 1, or 2 copy indicators across the $J \approx 4,000$ different SNPs, each multiplied by the externally estimated weights $w_{j'}$ that capture how strongly each SNP predicts education years.

\[ Z_i^\text{\textcolor{childColour}{Child}}
= \sum_{j = 1}^{J} w_j x_{i,j}^\text{\textcolor{childColour}{Child}}, \]
and similarly for child $i$'s parents, to give $Z_{m(i)}^\text{\textcolor{motherColour}{Mother}}$ and $Z_{f(i)}^\text{\textcolor{fatherColour}{Father}}$ from summing across binary values $x_{m(i),j}^\text{\textcolor{motherColour}{Mother}}$ and $x_{f(i),j}^\text{\textcolor{fatherColour}{Father}}$, respectively.

Because the transmitted variants are determined by random meiotic segregation conditional on parental genotypes, the inherited Ed PGI contains quasi-random within-family variation.
This means the Ed PGI can be decomposed into the component that was inherited as a result of random draws from parent(s) who had 1 copy of the variant, and the certainty component from draws where parent(s) had 0 or 2 copies.
\begin{equation}
    \label{eqn:pgi-child}
    Z_i^\text{\textcolor{childColour}{Child}}
    = \Egiven{Z_i^\text{\textcolor{childColour}{Child}}}{
        Z_{m(i)}^\text{\textcolor{motherColour}{Mother}},
            Z_{f(i)}^\text{\textcolor{fatherColour}{Father}}} + u_i,
\end{equation}
where $u_i$ is a mean zero random error term arising from Mendelian independent assortment.
The only determining factor for $u_i$ is the randomness of inheriting exact counts of SNPs $x_{i,j}^\text{\textcolor{childColour}{Child}}$ from their parents, relative to what they were expected to inherit, $\frac 12 \big( x_{m(i),j}^\text{\textcolor{motherColour}{Mother}} + x_{f(i),j}^\text{\textcolor{fatherColour}{Father}} \big)$.

Ideally, the expected inheritance is constructed from observed parental Ed PGI values, $Z_{m(i)}^\text{\textcolor{motherColour}{Mother}}, Z_{f(i)}^\text{\textcolor{fatherColour}{Father}}$.
However, parental PGI values are not observed for a large enough sample in the UKB for statistical inference.
Thus, I follow \cite{young2022mendelian} and leverage siblings' genetic data to impute parental PGI values.
The expected Ed PGI inheritance, therefore, is calculated using the same within-family variation that early genetic studies did by examining within family differences.
An individual's Ed PGI is drawn from the same distribution as their sibling, because they had the same parents; in absence of observing both parents directly, this approach uses the sibling to infer this distribution.

This approach is an acceptable substitute to the ideal case of using observed parents' data, because of the simple fact that children and parents are two independent draws from the same parental genetic distribution.
Write $Z_{s(i)}^\text{\textcolor{parentColour}{Sibling}}$ for $i$'s sibling's Ed PGI value, and $Z_{i, s(i)}^\text{\textcolor{parentColour}{Parental}}$ for the imputed parental value, calculated using genetic data from the individual and their observed sibling.
In expectation, a child and their observed sibling's Ed PGI together proxy for their unobserved parental PGI values.
\[ \E{Z_{i, s(i)}^\text{\textcolor{parentColour}{Parental}}}
    = \frac12 \Big(
        Z_{m(i)}^\text{\textcolor{motherColour}{Mother}}
            + Z_{f(i)}^\text{\textcolor{fatherColour}{Father}} \Big)
    = \Egiven{Z_i^\text{\textcolor{childColour}{Child}}}{
        Z_{m(i)}^\text{\textcolor{motherColour}{Mother}},
            Z_{f(i)}^\text{\textcolor{fatherColour}{Father}}}. \]

However, because siblings collectively reveal approximately 75\% of parental genetic variation, the resulting imputed values are measured with error relative to the true parental mean.
This reduces the precision with which the parental genetic component is controlled and may leave residual confounding in the estimated coefficient on the child's Ed PGI.
The resulting direction of bias is not generally determined by classical attenuation arguments because the mismeasured variable enters as a control rather than as the treatment of interest.

\begin{figure}[h!]
    \centering
    \singlespacing
    \caption{Distribution of the Ed PGI, Relative to Imputed Parental Decile.}
    \begin{subfigure}[b]{0.495\textwidth}
        \centering
        \caption{Ed PGI.}
        \includegraphics[width=\textwidth]{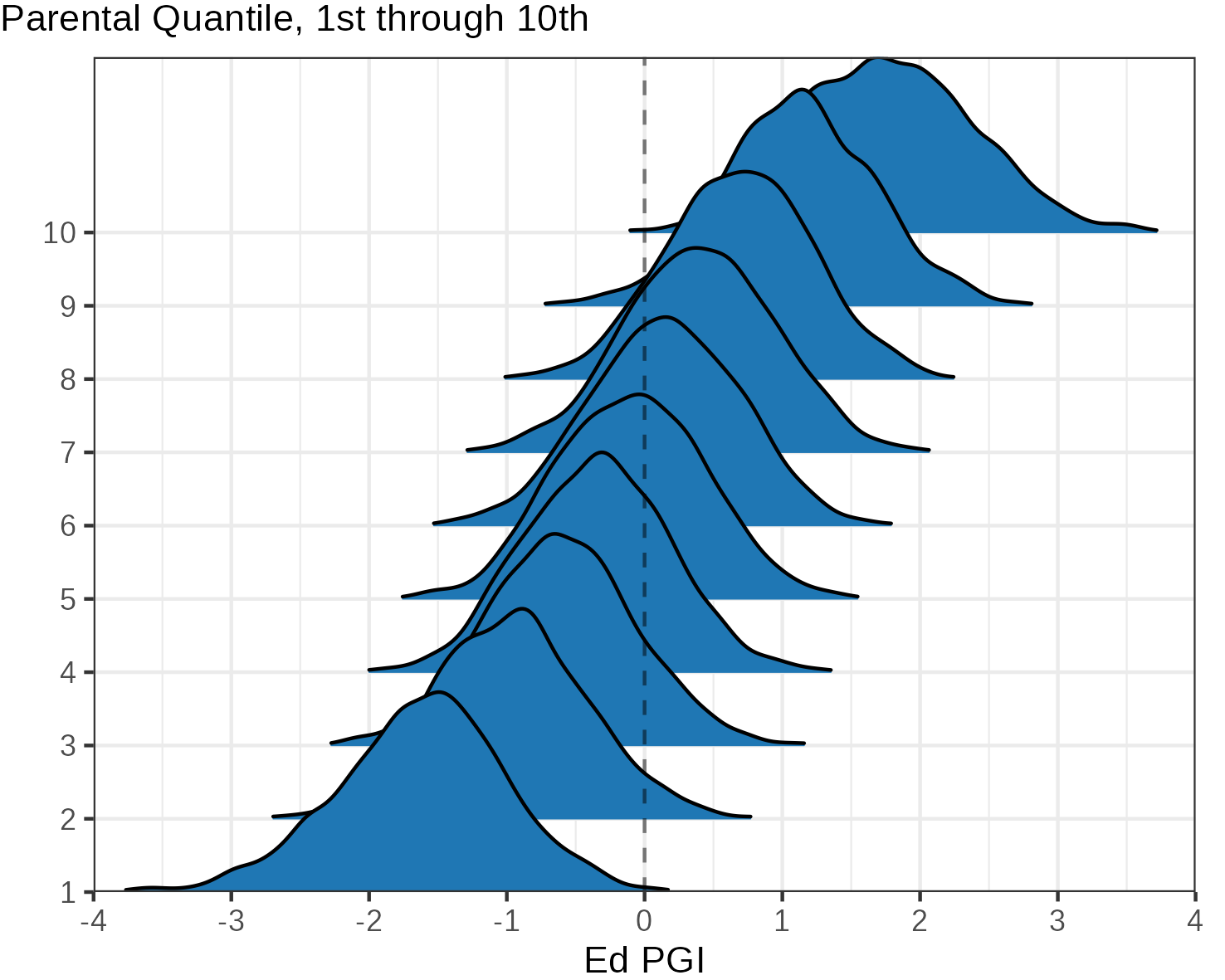}
        \label{fig:edpgi-self-dist}
    \end{subfigure}
    \begin{subfigure}[b]{0.495\textwidth}
        \centering
        \caption{Ed PGI minus imputed parental mean.}
        \includegraphics[width=\textwidth]{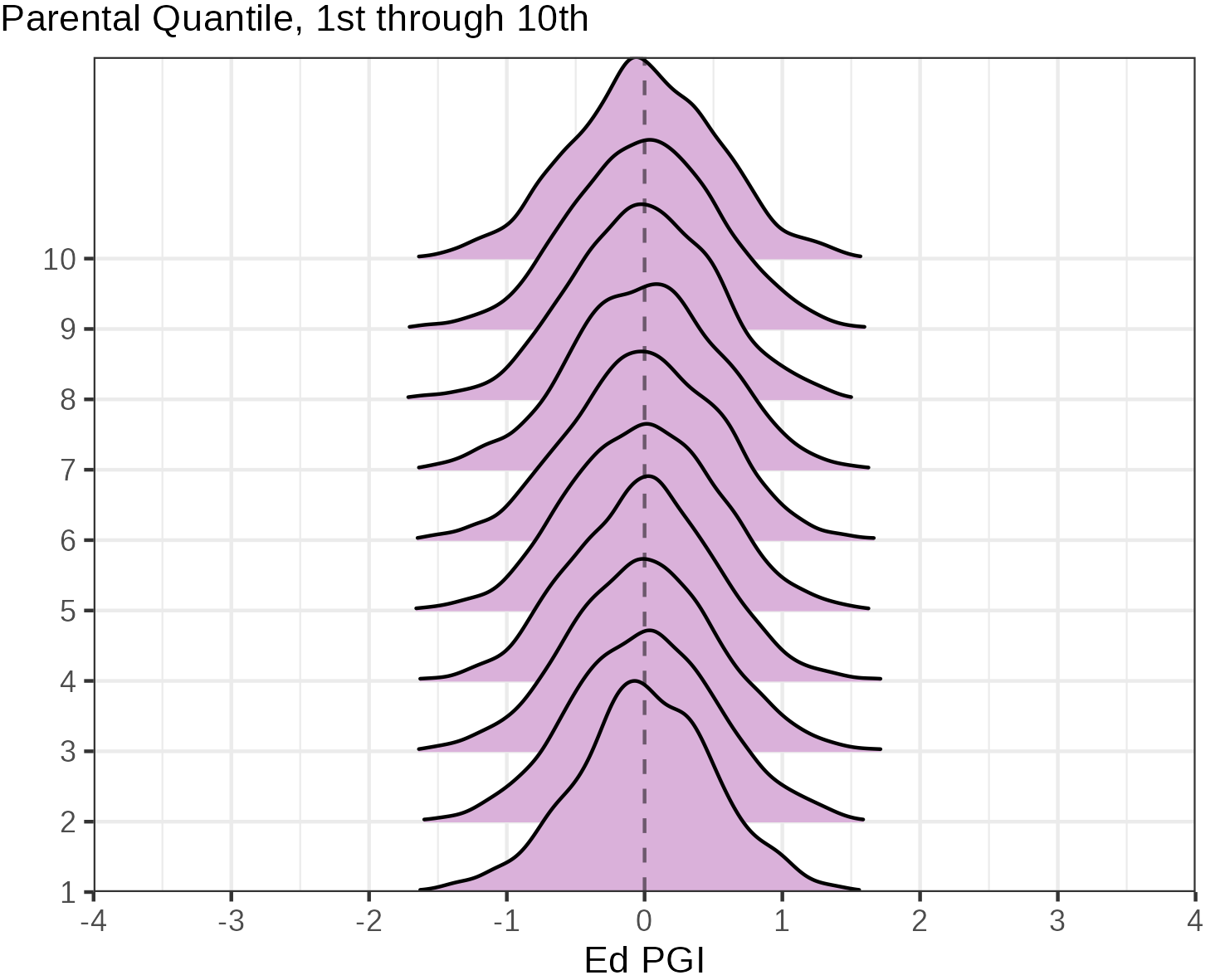}
        \label{fig:edpgi-random-dist}
    \end{subfigure}
    \label{fig:edpgi-dist}
    \vspace{-1cm}
    \justify
    \footnotesize
    \textbf{Note}:
    This figure shows the distribution of Ed PGI, $Z_i^\text{\textcolor{childColour}{Child}}$, and the Ed PGI minus imputed parental mean, $Z_i^\text{\textcolor{childColour}{Child}} - \Egiven{Z_i^\text{\textcolor{childColour}{Child}}}{
        Z_{m(i)}^\text{\textcolor{motherColour}{Mother}},
            Z_{f(i)}^\text{\textcolor{fatherColour}{Father}}} $, conditional on the decile of the parental imputed Ed PGI.
    Parental imputed Ed PGI decile 1 shows the distributions for the bottom 10\% of parental imputed values, decile 2 shows distribution among 10--20\%, and so on.
\end{figure}

Children with a higher Ed PGI value inherit this from their parents, and so had parents with higher Ed PGI values.
\autoref{fig:edpgi-dist} shows this tendency, where children with parents in the top decile of the imputed Ed PGI distribution have a mean Ed PGI 1.5 standard deviations above the mean --- and similarly for those at the bottom.\footnote{
    Panel A of \autoref{fig:edpgi-dist} replicates the tendency shown in \citet[Figure~1]{houmark2024nurture}, with a new data source.
}
This means parental Ed PGI values, with their effects on the children and the environment they are raised in, are a meaningful confounder for the effects of the child's own Ed PGI on education and labour market outcomes.
Panel B shows that, after conditioning for parental imputed values, there is no correlation across the distribution; this reflects the fact that Mendelian independent assortment induces quasi-random variation in the child's realised Ed PGI.

\subsection{Causal Identification Strategy}
Let $D_i(z)$ denote the potential education level and $Y_i(z, D_i(z))$ labour market outcome for individual $i$, under the hypothetical alternative realisation that $i$ had Ed PGI value $z$.
Observed education and wages/income data are equal to the potential outcomes for their realised Ed PGI value, $Z_i^\text{\textcolor{childColour}{Child}}$.
\[ D_i = D_i \left( Z_i^\text{\textcolor{childColour}{Child}} \right), \;\;\;\;
    Y_i = 
    Y_i \left( Z_i^\text{\textcolor{childColour}{Child}},
        D_i \left( Z_i^\text{\textcolor{childColour}{Child}} \right) \right).
\]
Potential labour market outcomes $Y_i(z, D_i(z))$ are a function of both the Ed PGI value, and resulting education $D_i(z)$.
This distinction is not important for measuring average genetic effects in this section, but will become the main focus in \autoref{sec:direct}.

\paragraph{Target Parameters.}
This study aims to estimate the average causal effect of having a higher Ed PGI, which is summarised by the following marginal average effects --- averaged across the distribution of $Z_i^\text{\textcolor{childColour}{Child}}$:\footnote{
    Throughout, I assume that both $D_i(z)$ and $Y_i(z, D_i(z))$ are differentiable for any $z$ across the distribution of $Z_i^\text{\textcolor{childColour}{Child}}$, together with standard regularity conditions permitting differentiation under the expectation.
}
\[ \text{Average education effect} \coloneq \E{\partialdiff{z} D_i(z)}, \;\;\;\; 
    \text{Average total effect} \coloneq
    \E{ \partialdiff{z} Y_i(z, D_i(z))}. \]
The average education effect is the average marginal change in education induced by a marginal increase in the Ed PGI;
the average total effect is the corresponding marginal change in labour-market income, allowing education to respond to the Ed PGI, \textit{ceteris paribus}.\footnote{
    Under the linear specification used below, these marginal effects are constant.
    Because the Ed PGI is standardised, the regression coefficients can therefore be interpreted as the effects of a one-standard-deviation increase in the Ed PGI.
}

\begin{figure}[h!]
    \centering
    \singlespacing
    \caption{Structural Causal Model for Ed PGI Effects.}
    \vskip0.25cm
    \label{fig:scm-edpgi}
    \begin{tikzpicture}
        \node[state,thick,ForestGreen] (mediator) at (0,0) {$D_i$};
        \node[state,thick,blue] (treatment) [left=5cm of mediator] {$Z_i$};
        \node[state,thick,red] (outcome) [below=1cm of mediator] {$Y_i$};
        \node[state,thick,blue] (treatment2) [left=5cm of outcome] {$Z_i$};
        \node[color=ForestGreen] [right=-0.01cm of mediator, align=left] {Education};
        \node[color=blue] [left=0.1cm of treatment, align=right] {Ed PGI};
        \node[color=blue] [left=0.1cm of treatment2, align=right] {Ed PGI};
        \node[color=red] [right=-0.01cm of outcome, align=left] {Occupation \\ Income $+$ Wages};
        \path[->, thick] (treatment) edge (mediator);
        \path[->, thick] (treatment2) edge (outcome);
        \node[color=orange] [align=center] at ($(treatment)!0.5!(mediator)$) {Average \\education effect};
        \node[color=orange] [align=center] at ($(treatment2)!0.5!(outcome)$) {Average \\total effect};
    \end{tikzpicture}
    \vskip0.25cm
    \justify
    \footnotesize
    \textbf{Note}:
    This figure shows the structural causal model, representing the two causal effects estimated in this section.
    The causal effects of the child's Ed PGI on education and later labour market outcomes are identified by controlling for parental (imputed) Ed PGI values.
\end{figure}
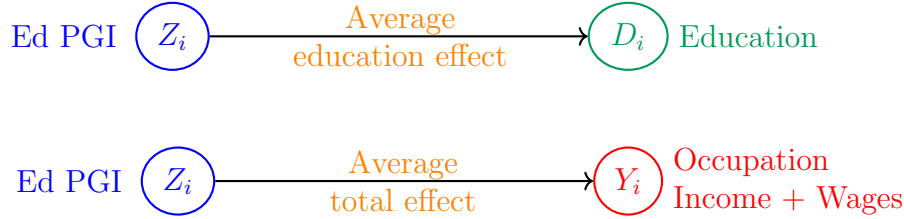

\paragraph{Identifying Assumptions.}
I rely on controlling for parental (imputed) Ed PGI values, for causal inference.
This yields estimates of causal effects if a child's Ed PGI is quasi-randomly assigned, conditional on their parents' Ed PGI mean --- the child's expected inheritance.
For any $z$ in the range of the Ed PGI distribution,
\begin{equation}
    Z_i^\text{\textcolor{childColour}{Child}}
    \indep \; D_i(z), \; Y_i(z, D_i(z))
    \;\;\;\; \Big | \;\;\;\; 
    Z_{i, s(i)}^\text{\textcolor{parentColour}{Parental}}.
\end{equation}

This identifying assumption is credible precisely because of the Mendelian independent variation described above.
The deviation away from expected inheritance represents the difference between a child's realised Ed PGI and the Ed PGI they were expected to inherit given their imputed parents' PGI values.
This deviation arises solely from the random segregation of chromosomes during meiosis: at each of the thousands of SNPs entering the Ed PGI, whether a parent with one copy passes it on is determined by an independent coin flip, with no connection to the family environment, the child's later circumstances, or any other factor that might influence education or labour market outcomes.
A child who drew a higher Ed PGI than their expected inheritance is therefore not systematically different in family background, parental environment, or any other confounding factor from one who drew a lower value; the only difference is the outcome of independent biological coin flips at conception.

\paragraph{Estimation Strategy.}
I estimate the average effects of Ed PGI values on education and labour market outcomes with the following regression equations.
\begin{align}
    \label{eqn:parametric-firststage}
    D_i &= \phi_1 + \pi Z_i^\text{\textcolor{childColour}{Child}}
        + \zeta_1 Z_{i, s(i)}^\text{\textcolor{parentColour}{Parental}}
        + \vec \varphi_1' \vec X_i
        + \eta_i,  \\
    \label{eqn:parametric-secondstage}
    Y_i &= \phi_2 + \theta Z_i^\text{\textcolor{childColour}{Child}}
        + \zeta_2 Z_{i, s(i)}^\text{\textcolor{parentColour}{Parental}}
        + \vec \varphi_2' \vec X_i
        + \epsilon_i,
\end{align}
where $\pi$ is the effect of the Ed PGI on education variables (average education effect), $\theta$ the effect on later labour market outcomes (average total effect).
Inclusion of parental imputed Ed PGI values yields causal inference, and $\vec X_i$ refers to a set of control variables including gender, sibling count, in city indicator, birth year and UKB data collection year fixed effects, and inclusion of other PGI values.\footnote{
    The other PGI variables are shown in Summary \autoref{tab:ukb-summary}, including estimated PGIs for ADHD, asthma, bipolar disorder, body mass index, diabetes (type 2), height, and schizophrenia.
}
Including these other PGI controls isolates the effects of the Ed PGI net of contribution from SNPs highly associated with other outcomes.

Using standard linear models, the marginal effects are assumed constant across the Ed PGI distribution, so that Ordinary Least Squares (OLS) consistently estimates a single parameter coinciding with the average total effect.
Without linearity, OLS recovers a variance-weighted average of marginal effects, placing disproportionate weight on individuals whose Ed PGI deviates most from its conditional mean; the linearity assumption is therefore what gives the estimates their interpretation as population averages.

Standard errors are clustered at the family unit level throughout.
The UKB recruited participants at the household level, which means that adult siblings are disproportionately represented among those who share a family unit --- either because they were co-resident at the time of recruitment or because the household-based design increased the probability that multiple siblings enrolled. This induces within-family correlation in the residuals that, if ignored, would cause conventional standard errors to understate sampling uncertainty.
Clustering at the family unit level accounts for arbitrary within-cluster correlation, yielding valid inference under the UKB sampling design.

\paragraph{Limitations.}
Several limitations apply to the estimates presented here.
First, the OLS estimates of average education effect and average total effects (\autoref{eqn:parametric-firststage} and \eqref{eqn:parametric-secondstage}) are not simple averages across individuals, but rather convex weighted averages of individual-level causal effects, with weights determined by the OLS projection \citep{angrist1999}.
Without assuming homogeneity of effects across the Ed PGI distribution, these estimates need not coincide with the population average total effect, and may place disproportionate weight on individuals whose Ed PGI deviates most from its conditional mean.

Second, the analysis sample is restricted to UKB participants of European ancestry with at least one observed family member, as described in \autoref{sec:data}.
Internal validity is sound within this sample, although the estimated average effects may not generalise to the broader British population, whose observed and unobserved structure may systematically differ from this selected group.

Third, the sibling-based imputation measures the parental PGI with error relative to observing both parental genotypes directly.
This may leave residual confounding in the estimated child-PGI effect, with a direction that is not determined in general.
Separately, estimation error in the GWAS weights used to construct the child's Ed PGI may attenuate estimates relative to those based on an infeasible oracle PGI.\footnote{
    The Ed PGI weights used in this paper (from \citealt{okbay2022polygenic}) are calculated with measurement error, leading to attenuation bias relative to an infeasible truth Ed PGI with oracle weights.
    See \cite{van2023overcoming} for further discussion on measurement error of this kind.
}

\section{Education and Total Genetic Effects}
\label{sec:genetic-effects}
The Ed PGI has substantial causal effects on both educational attainment and labour market outcomes.
\autoref{tab:genetic-effects} reports estimates of the average causal effects, identified by the quasi-random variation in genetic inheritance.
Each coefficient on the Ed PGI represents the average causal effect of a one standard deviation increase in the Ed PGI, conditional on the imputed parental Ed PGI and the full set of controls.
The parental Ed PGI enters as a control throughout: it is correlated with both the child's Ed PGI and the outcomes of interest, but does not represent a causal effect, as it absorbs the component of a child's Ed PGI that reflects inherited family environment rather than random genetic variation.

\begin{table}[!h]
    \singlespacing
    \centering
    \caption{Estimates of Education and Total Genetic Effects.}
    \small
    \centerline{
    \begin{tabular}{l c c c c c c}
        \\[-1.8ex]\hline \hline \\[-1.8ex] 
        & \multicolumn{3}{c}{Education Outcomes}
            & \multicolumn{3}{c}{Labour Market Outcomes} \\ \\
        Outcome:
        & Education & Age left & University & Occupation  & Occupation    & Household income  \\
        & years     & school & completion & hourly wage & annual income & (midpoint imputed)   \\
        \cmidrule(lr){2-4} \cmidrule(lr){5-7}
        & (1) & (2) & (3) & (4) & (5) & (6) \\
        \\[-1.8ex]\hline \\[-1.8ex]
 Ed PGI & 0.546 & 0.163 & 0.072 & 0.049 & 0.047 & 0.058 \\ 
   & (0.035) & (0.036) & (0.005) & (0.004) & (0.008) & (0.008) \\ 
  Parental Ed PGI & 0.870 & 0.337 & 0.105 & 0.067 & 0.066 & 0.094 \\ 
   & (0.049) & (0.048) & (0.007) & (0.006) & (0.010) & (0.011) \\ 
   \\[-1.8ex]\hline \\[-1.8ex] Outcome mean & 13.835 & 16.670 & 0.334 & 2.839 & 3.064 & 3.989 \\ 
  Adjusted $R^2$ & 0.146 & 0.043 & 0.107 & 0.138 & 0.205 & 0.136 \\ 
  Observations & 24,702 & 16,726 & 24,702 & 24,702 & 20,560 & 22,181 \\ 
  
        \\[-1.8ex]\hline \\[-1.8ex]
    \end{tabular}
    }
    \label{tab:genetic-effects}
    \justify
    \footnotesize
    \textbf{Note:}
    This table shows estimates of the average causal effect of the Ed PGI on education variables (average education effect) and labour market outcomes (average total effect) --- specification in \autoref{eqn:parametric-firststage} and \eqref{eqn:parametric-secondstage}.
    The coefficient on the Ed PGI represents the average causal effect, and coefficient on the parental Ed PGI is a control parameter.
    Specification for estimates in this table including gender, sibling count, in city indicator, birth year and UKB data collection year fixed effects, and inclusion of other PGI values as controls.
    Appendix~\autoref{tab:genetic-robust} shows estimates varying the control sets, and with sibling fixed effects.
\end{table}

The Ed PGI has a meaningful causal effect on education, with the largest effects operating through formal qualification thresholds.
A one standard deviation increase in the Ed PGI causes, on average, an additional 0.55 years of education (column 1; SE 0.04), against a sample mean of 13.84 years, and raises the probability of completing university by 7 percentage points on a baseline completion rate of 33 percent --- a relative increase of around one fifth (column 3).
Age of leaving full-time education, a common measure of education in the UK given that legal school-leaving age is defined by age rather than qualification, is observed for a subset of 16,726 participants; the estimated effect of 0.16 years (column 2; SE 0.04) is smaller than the qualification-based measure, reflecting that the Ed PGI operates more strongly through the decision to pursue higher qualifications than through the legally-determined school-leaving margin.

\begin{figure}[!h]
    \centering
    \singlespacing
    \caption{Effect of Ed PGI on Education and Wages.}
    \begin{subfigure}[b]{0.495\textwidth}
        \centering
        \caption{Ed PGI $\to$ Education Years.}
        \includegraphics[width=\textwidth]{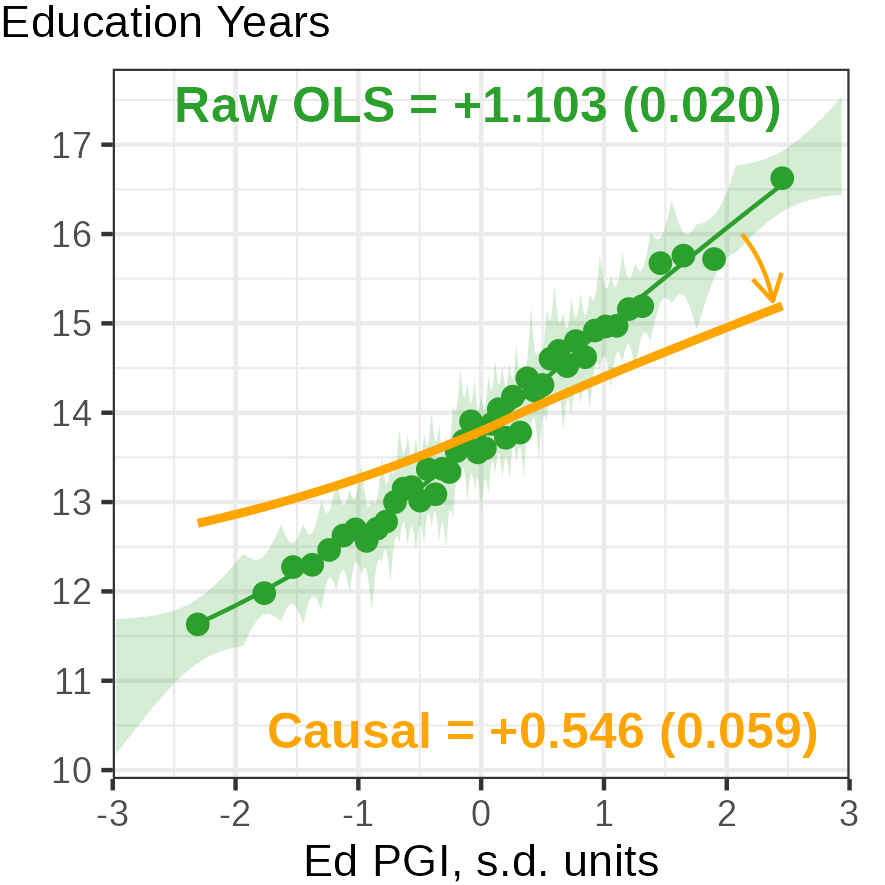}
        \label{fig:causal-edpgi-edyears}
    \end{subfigure}
    \begin{subfigure}[b]{0.495\textwidth}
        \centering
        \caption{Ed PGI $\to$ Hourly Wages.}
        \includegraphics[width=\textwidth]{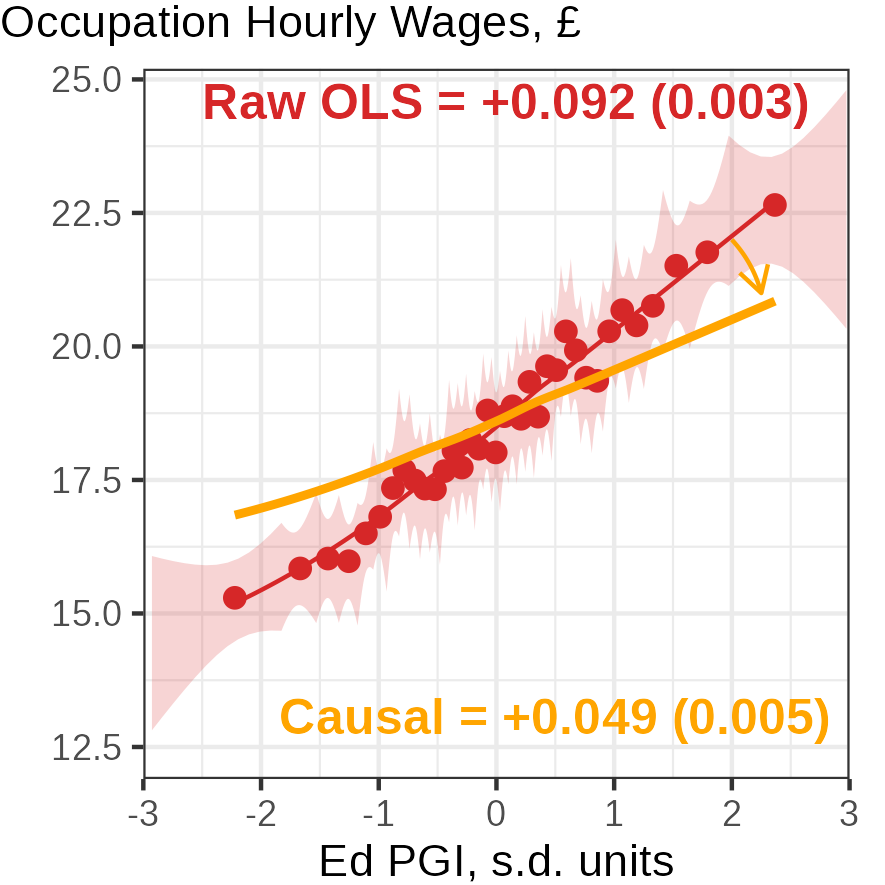}
        \label{fig:causal-edpgi-earnings}
    \end{subfigure}
    \label{fig:causal-edpgi-effects}
    \vspace{-1cm}
    \justify
    \footnotesize
    \textbf{Note}:
    This figure shows the correlation between Ed PGI and education years, as well as occupation coded hourly wages.
    The shifted lines are the estimates of the causal effect, estimated by controlling for parental imputed Ed PGI values.
\end{figure}

The Ed PGI has a consistent and precisely estimated causal effect on earnings across all three income measures.
A one standard deviation increase in the Ed PGI raises hourly wages by approximately 5 percent (column 4) and annual occupation income by 5 percent (column 5), with a slightly larger effect of 6 percent on household income (column 6).
These estimates capture the total average causal effect of the Ed PGI on labour market outcomes --- combining any direct biological pathway with effects mediated through the additional education documented above.

\begin{table}[!h]
    \singlespacing
    \centering\newpage
    \caption{Different Specifications for Estimates of Average Education and Total Effects.}
    \small
    \begin{tabular}{l c c c c c c}
        \\[-1.8ex]\hline \hline \\[-1.8ex] 
        & \multicolumn{3}{c}{Education Outcomes}
            & \multicolumn{3}{c}{Labour Market Outcomes} \\ \\
        Outcome:
        & Education & Age left & University & Occupation  & Occupation    & Household \\
        & years     & school & completion & hourly wage & annual income & income    \\
        \cmidrule(lr){2-4} \cmidrule(lr){5-7}
        & (1) & (2) & (3) & (4) & (5) & (6) \\
        \\[-1.8ex]\hline \\[-1.8ex]
        \multicolumn{7}{l}{\textbf{Panel A:}
            Raw Correlation} \\
 Ed PGI & 1.103 & 0.361 & 0.140 & 0.092 & 0.081 & 0.109 \\ 
   & (0.020) & (0.020) & (0.003) & (0.003) & (0.005) & (0.005) \\ 
  
        \\[-1.8ex]\hline \\[-1.8ex]
        \multicolumn{7}{l}{\textbf{Panel B:}
            Including Parental Ed PGI with no Control Variables} \\
 Ed PGI & 0.565 & 0.161 & 0.074 & 0.050 & 0.042 & 0.056 \\ 
   & (0.035) & (0.036) & (0.005) & (0.005) & (0.009) & (0.008) \\ 
  Parental Ed PGI & 0.855 & 0.327 & 0.105 & 0.067 & 0.062 & 0.084 \\ 
   & (0.049) & (0.049) & (0.007) & (0.006) & (0.012) & (0.012) \\ 
  
        \\[-1.8ex]\hline \\[-1.8ex]
        \multicolumn{7}{l}{\textbf{Panel C:}
            Including Family Fixed Effects} \\
 Ed PGI & 0.560 & 0.294 & 0.070 & 0.052 & 0.061 & 0.065 \\ 
   & (0.039) & (0.053) & (0.006) & (0.005) & (0.010) & (0.010) \\ 
   \\[-1.8ex]\hline \\[-1.8ex] Outcome mean & 13.835 & 16.670 & 0.334 & 2.839 & 3.064 & 3.989 \\ 
  Observations & 24,702 & 16,726 & 24,702 & 24,702 & 20,560 & 22,181 \\ 
  
        \\[-1.8ex]\hline \\[-1.8ex]
    \end{tabular}
    \label{tab:genetic-robust}
    \justify
    \footnotesize
    \textbf{Note:}
    This table shows estimates of the average causal effect of the Ed PGI on education variables (average education effect) and labour market outcomes (average total effect) --- specification in \autoref{eqn:parametric-firststage} and \eqref{eqn:parametric-secondstage}.
    The coefficient on the Ed PGI represents the average causal effect, and coefficient on the parental Ed PGI is a control parameter.
    Specifications vary for each Panel, as described.
\end{table}

The main estimates are robust to alternative specifications, and the main causal estimates are half as large as  raw correlation between the Ed PGI for all outcomes under study.
\autoref{tab:genetic-robust} reports estimates under alternative 
specifications, and this raw correlation.
Panel A shows the raw correlations between the Ed PGI and outcomes, without 
conditioning on parental PGI or any controls; these are roughly twice the size of the causal estimates, illustrating the extent of confounding when parental genetics are not accounted for.
Panel B re-estimates the main specification without the additional control variables, and finds estimates nearly identical to those in \autoref{tab:genetic-effects}, confirming that the causal identification rests on the parental PGI rather than the auxiliary controls.
Including the other PGI for other outcomes as controls make little difference to the causal estimates, which aligns with specifications reporting correlations with the Ed PGI that do attenuate with inclusion of other PGIs (e.g., \citealt{bolyard2025understanding}).
Panel C replaces the parental PGI with sibling fixed effects, which partial out all shared family characteristics directly; the resulting estimates are again very close to the main results across education and labour market outcomes, lending further support to the identifying assumption that within-family variation in the Ed PGI is quasi-randomly assigned.

These findings closely replicate the main empirical specification and results in \citet{carvalho2024genetics}.
The Ed PGI has meaningful causal effects on education and labour-market outcomes, while conditioning on the imputed parental Ed PGI substantially reduces the raw associations. The purpose of this replication is to establish the two inputs required for the analysis that follows: an Ed PGI effect of approximately 0.55 years on education, and a total income effect of approximately 5 percent.
Having established these inputs, I now turn to the paper's main contribution: decomposing the total effect of the Ed PGI on earnings into the share operating through education and a residual direct effect.

\section{Direct and Indirect Effects}
\label{sec:direct}
The average total effect of the Ed PGI on labour market outcomes is the sum of two conceptually distinct channels.
The first is a direct channel: genetic predisposition toward education may influence earnings independently of education itself, through traits such as cognitive ability, non-cognitive skills, or personality characteristics that are correlated with the Ed PGI and valued by employers.
The second is an indirect channel: a higher Ed PGI induces more education, and that additional education generates a labour market return.

The size and robustness of the labour market effects documented in 
\autoref{sec:genetic-effects} has led to a common interpretation in the literature that at least part of the Ed PGI's influence must operate through the direct channel.
This view is often motivated by the fact that the Ed PGI predicts scores on tests aiming to measure fluid intelligence \citep{carvalho2024genetics}, by results showing earnings associations that persist even after conditioning on education \citep{papageorge2020genes,bryson2025gender}, and in work arguing that the Ed PGI captures skills related to financial literacy and other non-educational attributes \citep{barth2020genetic}.
Separating these channels matters for understanding the biological and social mechanisms behind genetic effects on earnings.

This section uses a Causal Mediation (CM) framework to decompose the average total effect of the Ed PGI into its direct and indirect components, where years of education serves as the mediating variable.
The key empirical challenge is that causal mediation analysis requires not only that the Ed PGI is quasi-randomly assigned (conditional on parental values), but also that the mediator is quasi-randomly assigned conditional on treatment, which is not credible for education in this setting.\footnote{
    \appendixref{appendix:iv} expands on this point, connecting to the instrumental variables literature.
}
UKB participants were not randomly assigned education, so it takes further steps to identify the returns to an additional year of education.
Based on this insight, I combine credible estimates of the average effect of the Ed PGI on education together with estimates of the returns to education to identify the direct and indirect effect channels.

\subsection{Causal Mediation (CM) Framework}
\label{sec:cm-framework}
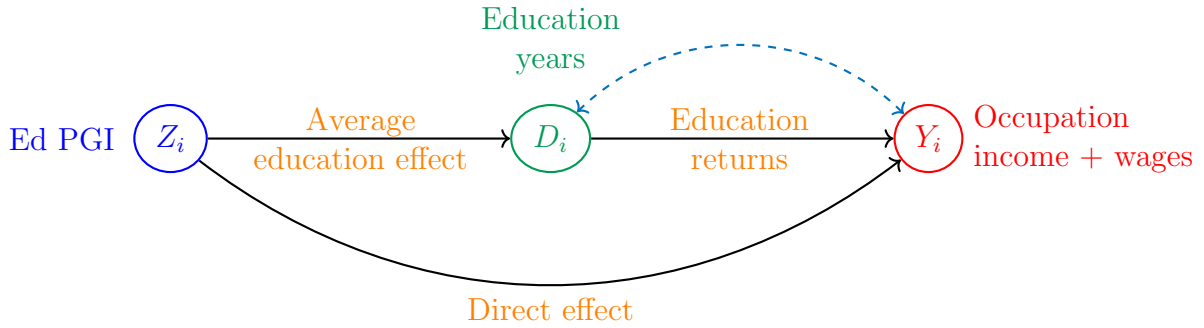
\begin{figure}[!h]
    \centering
    \singlespacing
    \caption{Structural Causal Model for Ed PGI Effects.}
    \label{fig:scm-cm}
    \begin{tikzpicture}
        \node[state,thick,ForestGreen] (mediator) at (0,0) {$D_i$};
        \node[state,thick,blue] (treatment) [left=4cm of mediator] {$Z_i$};
        \node[state,thick,red] (outcome) [right=4cm of mediator] {$Y_i$};
        \node[color=ForestGreen, align=center] [above=0.25cm of mediator] {Education \\ years};
        \node[color=blue] [left=0.1cm of treatment, align=right] {Ed PGI};
        \node[color=red] [right=-0.01cm of outcome, align=left] {Occupation \\ income $+$ wages};
        \path[->, thick] (treatment) edge (mediator);
        \path[->, thick] (mediator) edge (outcome);
        \path[->, thick] (treatment) edge[bend right=37.5] (outcome);
        \node[color=orange] [below=1.5cm of mediator] {Direct effect};
        \node[color=orange] [align=center] at ($(treatment)!0.5!(mediator)$) {Average \\education effect};
        \node[color=orange] [align=center] at ($(mediator)!0.5!(outcome)$) {Education \\ returns};
        \path[<->,dashed,thick,color=RoyalBlue] (mediator) edge[bend right=-45] (outcome);
    \end{tikzpicture}
    \justify
    \footnotesize
    \textbf{Note}:
    This figure shows the structural causal model behind direct and indirect effects, where $Z_i \to D_i$ represents a causal relationship of $Z_i$ on $D_i$.
    The dotted line between education and income \& wages represents the fact that education is not randomly assigned, and a strategy for identifying returns to education is needed for valid CM estimates.
\end{figure}

\paragraph{Target Parameters.}
The average total effect of the Ed PGI on labour market outcomes can be 
decomposed into a direct and indirect component.
Formally, the direct and indirect effects are defined as:
\begin{align*}
    \text{Direct Effect} &=
        \E{\partialdiff{z} Y_i(z, D_i(Z_i^\text{\textcolor{childColour}{Child}}))}, \\
    \text{Indirect Effect} &=
        \E{\partialdiff{z}
            Y_i(Z_i^\text{\textcolor{childColour}{Child}}, D_i(z))}.
\end{align*}
The direct effect captures the effect of the Ed PGI on earnings holding the education channel fixed; the indirect effect captures the effect that operates exclusively through the induced change in education.
These sum to the average total effect $\theta$ estimated in 
\autoref{sec:genetic-effects},
\begin{equation*}
    \label{eqn:ape-decomp}
    \underbrace{
        \E{\partialdiff{z} Y_i(z, D_i(z))}
    }_{\text{Average total effect, } \theta}
    =
    \underbrace{
        \E{\partialdiff{z} 
        Y_i(z, D_i(Z_i^\text{\textcolor{childColour}{Child}}))}
    }_{\text{Direct effect}}
    +
    \underbrace{
        \E{\partialdiff{z} 
        Y_i(Z_i^\text{\textcolor{childColour}{Child}}, D_i(z))}
    }_{\text{Indirect effect}}.
\end{equation*}
This decomposition holds pointwise and is then averaged over individuals, with each effect evaluated at the realised Ed PGI value,
$z = Z_i^\text{\textcolor{childColour}{Child}}$; the linear specification assumes that these average marginal effects do not vary across the $Z_i^\text{\textcolor{childColour}{Child}}$ distribution.
\autoref{fig:scm-cm} illustrates this decomposition in the structural causal model: the direct effect corresponds to the Ed PGI arrow bypassing education, and the indirect effect to the path running through education years and its labour market return.\footnote{
    Throughout, ``direct'' means direct with respect to completed education years: it is the residual channel not operating through this measured mediator, rather than a purely biological effect.
}

\paragraph{Identification.}
Non-parametric CM identification requires that the mediator is quasi-randomly assigned conditional on treatment \citep[see][]{imai2010identification}.
This assumption is not credible here: education years are shaped by family background, ability, and a host of other factors that independently influence earnings, so that the education--earnings relationship might not represent the average causal effect for mediator compliers in the UKB sample (even after conditioning on the Ed PGI).
However, it is possible to parametrically identify the direct and indirect effects with a valid estimate for the causal returns to education, by the chain rule.\footnote{
    This additionally requires the smoothness and regularity conditions assumed in \autoref{sec:genetics}.
    I also assume that the return to education relevant for the education margin shifted by the Ed PGI can be represented as separable from the average education effect; \autoref{appendix:mediation-identification} discusses this assumption in detail.
}
\begin{equation*}
    \label{eqn:indirect-chain}
    \underbrace{
        \E{\partialdiff{z} Y_i(Z_i^\text{\textcolor{childColour}{Child}}, D_i(z))}}_{
            \text{Indirect Effect}}
    = \underbrace{\E{ \partialdiff{z} D_i(z)}}_{\text{Average education effect, } \pi}
    \times \;\;\;\;
    \underbrace{\E{\partialdiff{d} Y_i(Z_i^\text{\textcolor{childColour}{Child}}, d)}}_{\text{Returns to education, } 
    \beta}.
\end{equation*}

The indirect effect equals the average education effect $\pi$ --- credibly 
identified by the Mendelian design --- multiplied by the returns to an additional year of education, $\beta$.
With heterogeneous treatment effects, $\beta$ represents the average education returns for people pushed into more years of education.\footnote{
    Note that this is a parametrically identified model, because the correlational measure (and accompanying sensitivity analysis) is meant to inform the local average effect for $D_i(Z_i^\text{\textcolor{childColour}{Child}})$ compliers --- but does not non-parametrically identify it.
}
The direct effect then follows as average total effect minus the indirect effect,
\begin{equation*}
    \label{eqn:direct-residual}
    \text{Direct Effect} = \theta - \pi\beta.
\end{equation*}

To interpret $\beta$ as an average return to education for individuals whose schooling is shifted by the Ed PGI, I additionally assume that no individual has a negative education response, $\partial D_i(z)/\partial z \geq 0$.
The chain-rule decomposition remains valid without this monotonicity condition, but the implied average return may then place negative weight on some individual education returns \citep{blackwell2024assumption}.

\paragraph{Estimation.}
I estimate the mediation second-stage from the following specification:
\begin{equation}
    \label{eqn:ed-returns}
    Y_i = \alpha + \beta D_i + \gamma Z_i^\text{\textcolor{childColour}{Child}} 
        + \delta Z_i^\text{\textcolor{childColour}{Child}} D_i 
        + \zeta Z_{i, s(i)}^\text{\textcolor{parentColour}{Parental}}
        + \vec\varphi' \vec X_i + \varepsilon_i,
\end{equation}
where the partial return to education is $\beta+\delta Z_i^\text{\textcolor{childColour}{Child}}$, and the partial direct effect of the Ed PGI is $\gamma+\delta D_i$.
The parameters $\beta$ and $\gamma$ are the corresponding effects when the interacted variable equals zero, while $\delta$ allows both effects to vary with the other variable.
Empirically, the estimated interaction is close to zero, so that the average return to education is approximately $\beta$ and the average direct effect is approximately $\gamma$.\footnote{
    \cite{mediation-natural-experiment} shows that the interaction term $\delta$ does not suffer from selection bias even when the mediator is not randomly assigned, provided the treatment is quasi-randomly assigned --- a condition satisfied here by the Mendelian design.
    So that near-zero estimates of the interaction term in \autoref{tab:ukb-ed-returns} mean the interaction term is irrelevant.
}

SEs for the mediation estimates are obtained via a cluster bootstrap, resampling family units rather than individual observations.
This preserves the within-family correlation in the residuals arising from the UKB sampling design, while also accommodating the non-linearity of the indirect effect estimator --- the product of two estimated quantities.
Each bootstrap draw resamples family units with replacement, estimates the mediation first- and second-stage parameters, and computes the implied direct and indirect effects; SEs are the standard deviations of the resulting bootstrap distributions.

\paragraph{Limitations.}
Several limitations apply to the CM estimates presented in this section.
First, the OLS estimates of the average total effect and average education effect are variance-weighted averages of individual-level effects under heterogeneity (as noted in \autoref{sec:genetics}).
Without the linearity assumption, these average effects need not coincide with population average effects.

Second, and more substantively, the indirect effect requires the supplied education return estimates to represent the return relevant for the education margins shifted by the Ed PGI.
One sufficient condition is homogeneous returns to education; a weaker sufficient condition permits heterogeneous returns, but requires them to be uncorrelated with individuals' education responses to the Ed PGI.
More generally, the validity of the mediation estimates rests on the credibility of approximating the mediator-relevant returns to education (discussed further in \autoref{appendix:mediation-identification}).
The only remaining step for identifying CM effects is to provide credible estimates of the causal returns to education in the UKB sample.

\subsection{Returns to Education}
Correlational estimates of education returns are the most natural starting point.
A large literature finds that OLS estimates of the return to an additional year of schooling are remarkably stable; they are typically estimated in the range of 5--10 percent in log earnings, and close to estimates from quasi-experimental designs \citep{angrist1991does, harmon1995estimates, oreopoulos2006estimating}.
Two forces likely explain this convergence: ability bias pushes OLS upward, while measurement error in reported schooling pushes it downward, and the two roughly cancel. Quasi-experimental designs, meanwhile, often identify effects among specific groups of marginal compliers, whose average education returns need not equal the population average \citep{buscha2015wage}.
The fact that OLS and quasi-experimental estimates land in a similar range is informative, not coincidental.

\begin{table}[!h]
    \small
    \singlespacing
    \centering
    \caption{Returns to Education, Correlational Estimates.}
    \centerline{
    \begin{tabular}{l c c c c c c c c c c }
        \\[-1.8ex]\hline \hline \\[-1.8ex]
        Outcome: &
        \multicolumn{3}{c}{Occupation} & \multicolumn{3}{c}{Occupation} & \multicolumn{3}{c}{Household income} \\ 
        & \multicolumn{3}{c}{hourly wage} & \multicolumn{3}{c}{annual income} & \multicolumn{3}{c}{(midpoint imputed)} \\
        \cmidrule(lr){2-4} \cmidrule(lr){5-7} \cmidrule(lr){8-10}
        & (1) & (2) & (3) & (4) & (5) & (6) & (7) & (8) & (9) \\
        \\[-1.8ex]\hline \\[-1.8ex]
 Education years & 0.061 & 0.059 & 0.059 & 0.061 & 0.059 & 0.059 & 0.068 & 0.063 & 0.063 \\ 
   & (0.001) & (0.001) & (0.001) & (0.001) & (0.001) & (0.001) & (0.002) & (0.002) & (0.002) \\ 
  Ed PGI &  & 0.010 & 0.000 &  & 0.008 & -0.001 &  & 0.008 & -0.015 \\ 
   &  & (0.011) & (0.010) &  & (0.021) & (0.021) &  & (0.013) & (0.010) \\ 
  Education years $\times$ Ed PGI &  & 0.001 & 0.001 &  & 0.001 & 0.001 &  & 0.003 & 0.003 \\ 
   &  & (0.001) & (0.001) &  & (0.001) & (0.001) &  & (0.001) & (0.001) \\ 
  Parental Ed PGI &  &  & 0.016 &  &  & 0.016 &  &  & 0.040 \\ 
   &  &  & (0.003) &  &  & (0.005) &  &  & (0.007) \\ 
  \midrule Collected Education Returns & 0.061 & 0.059 & 0.059 & 0.061 & 0.059 & 0.059 & 0.068 & 0.064 & 0.063 \\ 
   & (0.001) & (0.001) & (0.001) & (0.001) & (0.001) & (0.001) & (0.002) & (0.002) & (0.002) \\ 
  \midrule $R^2$ & 0.328 & 0.332 & 0.332 & 0.273 & 0.274 & 0.274 & 0.204 & 0.208 & 0.208 \\ 
  Observation count & 24,702 & 24,702 & 24,702 & 20,560 & 20,560 & 20,560 & 22,181 & 22,181 & 22,181 \\ 
  
        \\[-1.8ex]\hline \\[-1.8ex]
    \end{tabular}
    }
    \label{tab:ukb-ed-returns}
    \justify
    \footnotesize
    \textbf{Note:}
    This table shows simple OLS estimates for returns to education in the UK Biobank sibling sample --- the relevant sample for causal genetic effects.
    All outcomes are in a log specification, and each column is a regression including the controls previously mentioned, with specification in \autoref{eqn:ed-returns}, the inclusion of the Ed PGI and parental Ed PGI as indicated.
    Standard errors are clustered at the family level.
    Observation counts vary because each outcome has a different number of observations which are non-missing.
    The final row for education returns collects the partial effect of education years, including the interaction terms between education years and the Ed PGI.
    Point estimates are relatively similar to those using the MSLA rise as an instrument (see \appendixref{appendix:MSLA}), with much more precision.
\end{table}

The correlational estimate for the labour market return to an additional year of education in the UKB analysis sample sits consistently at around 6 percent in log earnings across all three income measures (shown in \autoref{tab:ukb-ed-returns}).
This estimate is stable across specifications: adding controls for the Ed PGI and its interaction with education years barely moves the point estimates, and including the parental Ed PGI changes nothing.
The UKB correlational estimates are, in short, consistent with the broader literature.

Canonical labour economics work was concerned that the correlation between years of completed education and later-life income would be confounded by unobserved ability \citep{griliches1977}.
One may expect that genetic variants, contained in the Ed PGI, would be correlated with unobserved ability, and thus controlling for the Ed PGI may lead to lower correlational estimates for the returns to education.
This is not the case in UKB data, and is consistent with follow-up research that uses causal methods such as instrumental variables and finds estimates for returns to education similar to the correlational evidence (see \citealt{card1999causal} for an overview).
The stability of the return to education after conditioning on the Ed PGI and parental Ed PGI therefore suggests that ability bias, at least insofar as it operates through genetic endowment, is not a material source of upward distortion in the correlational estimate.

Nonetheless, it would still not be credible to assume years of education are randomly assigned, so that no correlational estimate is entirely credible on its own.
The UKB does contain a plausible quasi-experiment: the 1972 rise in the British Minimum School Leaving Age (MSLA), which compelled children born after September 1957 to stay in school until age 16.
While this sample lines up well with the UKB population, the resulting fuzzy RDD estimates of education returns (in the range of 4--6 percent in log earnings) carry standard errors too large to rule out either zero or substantially larger effects, and are therefore too imprecise for use in the mediation analysis.\footnote{
    \appendixref{appendix:MSLA} details estimates for labour market education returns using the MSLA rise as a regression discontinuity, as previously considered by \cite{barcellos2021effect}.
}
Given this, the correlational estimates anchored in the broader UK literature offer a more useful basis for the parametric mediation analysis, and the sensitivity analysis below makes the dependence on any particular return value transparent.

\subsection{Sensitivity Analysis}
The indirect effect is the product of two quantities: the effect on education years and the return to an additional year of education. Since these multiply together, any uncertainty in the education return translates directly and proportionally into uncertainty about the share of the total genetic effect that operates through education.
Rather than conditioning the mediation analysis on a single return value, I vary it across its plausible range.
In addition, this range is defined not arbitrarily, but based on the previous labour economics on returns to education.

\begin{figure}[!h]
    \centering
    \singlespacing
    \caption{Distribution of Economics Literature for Income Returns to Education.}
    \begin{subfigure}[b]{0.495\textwidth}
        \centering
        \includegraphics[width=\textwidth]{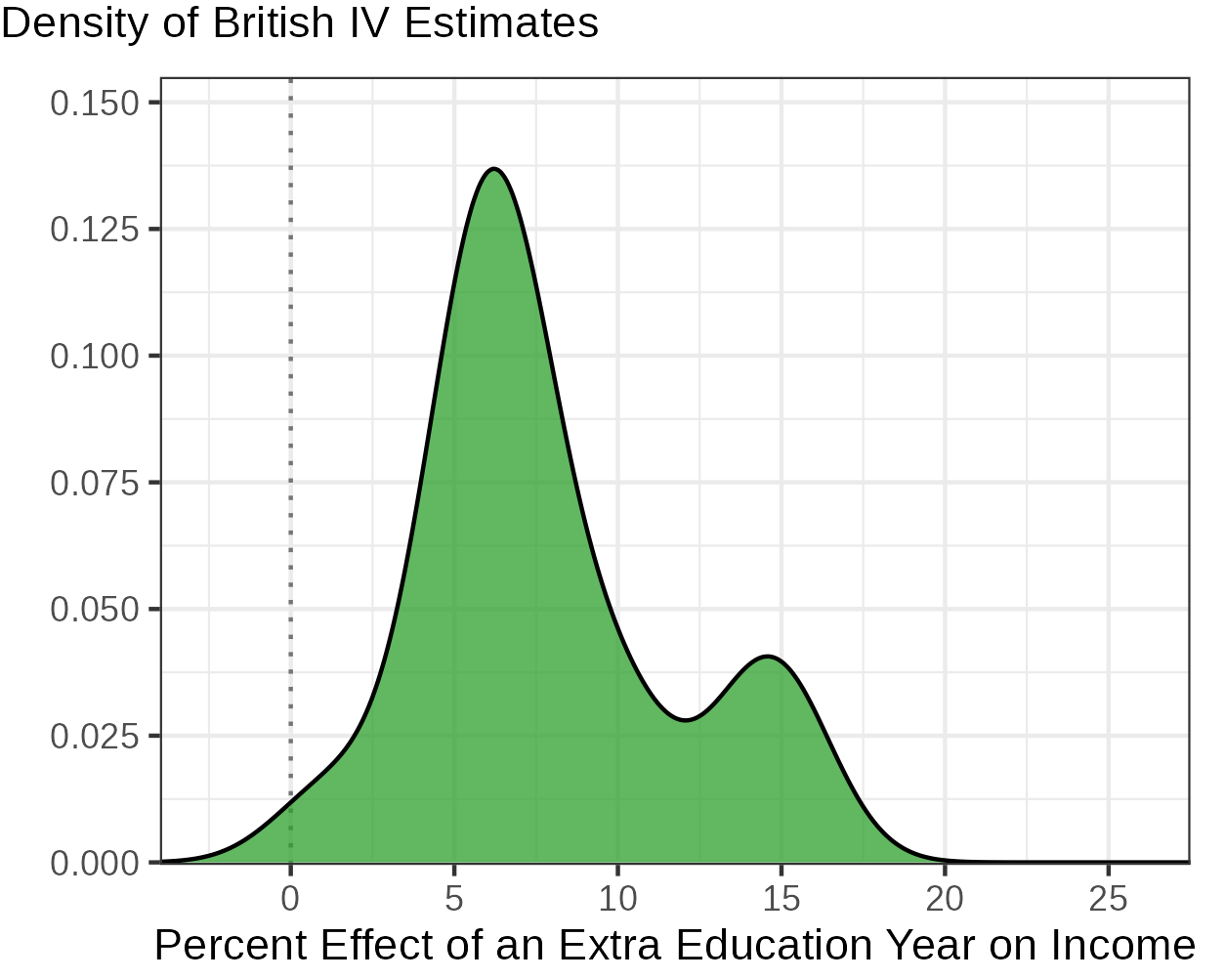}
    \end{subfigure}
    \begin{subfigure}[b]{0.495\textwidth}
        \centering
        \includegraphics[width=\textwidth]{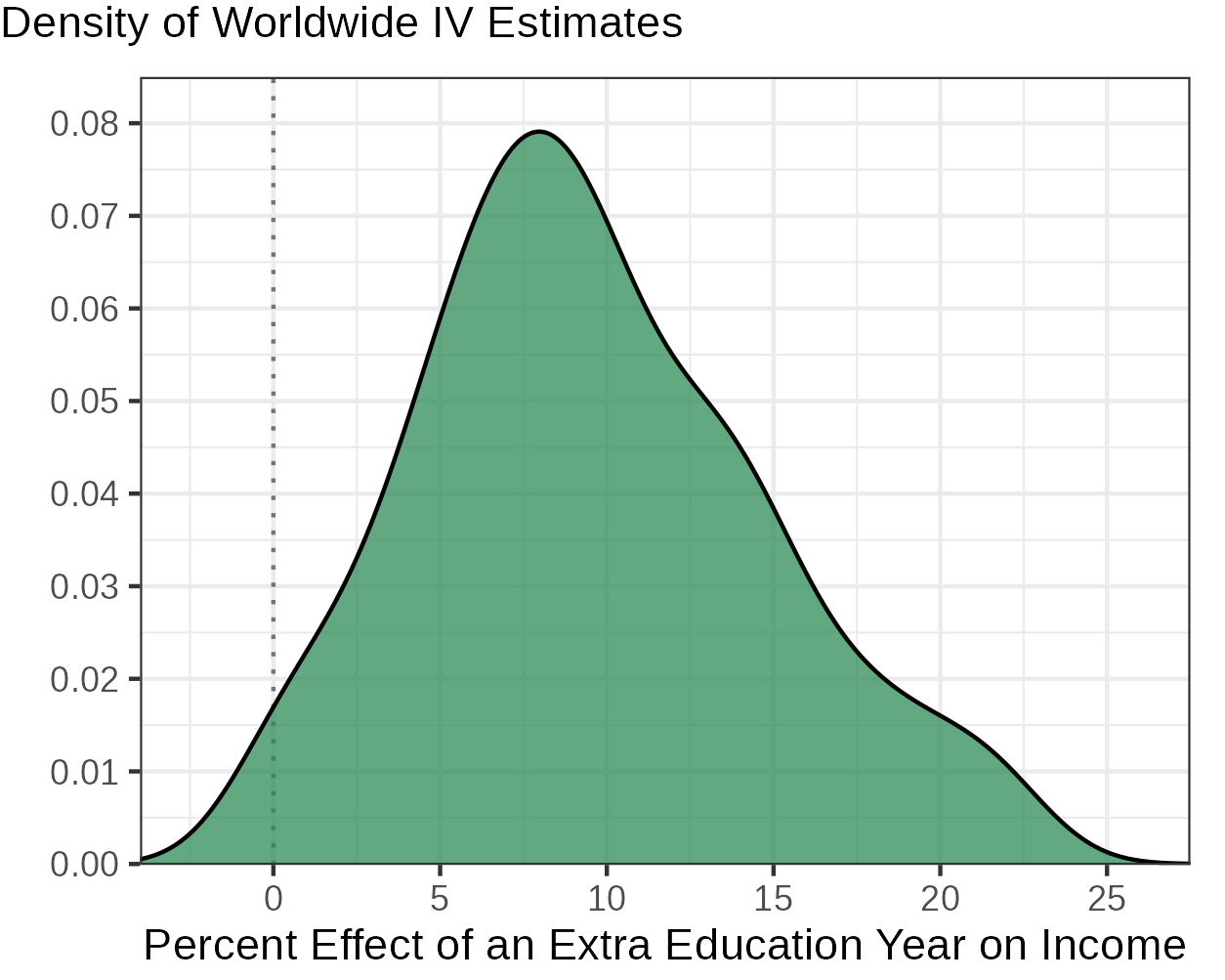}
    \end{subfigure}
    \label{fig:edreturns}
    \vspace{-0.5cm}
    \justify
    \footnotesize
    \textbf{Note}:
    These figures show the estimated kernel density for the distribution of estimates for later-life income returns to education.
    Panel A refers to the implied distribution from 17 estimates using IVs in the UK (shown in full in Appendix \autoref{tab:uk-meta}), and Panel B for 182 estimates from 144 studies across the world.
    These estimates were collected by \cite{patrinos2025causal}.
\end{figure}

The returns to an extra year of education in the UK has been estimated many times in the economics literature, giving a rich empirical base for calibrating the sensitivity analysis.
\autoref{fig:edreturns} shows the distribution of 17 quasi-experimental UK estimates collected by \cite{patrinos2025causal}, listed in full in Appendix~\autoref{tab:uk-meta}.
The distribution is right-skewed: most estimates fall between 4 and 8 percent, consistent with the correlational estimates in \autoref{tab:ukb-ed-returns}, but a tail of larger estimates reaches to around 15 percent.
These higher values tend to come from studies exploiting the 1947 compulsory schooling reform, which identifies returns among a different and older complier population than the UKB cohort.
The worldwide distribution from 182 estimates across 144 studies, shown alongside, is broadly consistent with the UK evidence.

The first step of the sensitivity analysis asks a deliberately agnostic question: how do the mediation estimates change as the assumed education return varies across the full span of values that appear in the literature, from 0 to 20 percent?
For each value in this range, the indirect effect follows directly as the product of the estimated genetic education effect and the assumed return, with the direct effect as the remainder.
This step takes no stand on which return value is correct, and shows whether the mediation conclusions are stable or fragile across the plausible range.

The second step uses the empirical UK distribution more directly.
I bootstrap across the 17 IV estimates in the economic literature (shown in Appendix~\autoref{tab:uk-meta}), drawing a return value at each replication and computing the implied mediation decomposition.
Repeating this across many draws produces a distribution of indirect and direct effect estimates that reflects both sampling uncertainty in the effect of the Ed PGI on education years and the genuine disagreement in the literature about the magnitude of education returns.
Reporting the point estimate, and bootstrapped standard errors of this distribution gives an empirical sensitivity distribution based on the collected UK estimates, rather than treating all values across the plausible range as equally likely.

Together, the two steps answer different questions. The first shows the full range of possible conclusions. The second shows where the weight of evidence places them.

\section{Direct and Indirect Effect Estimates}
\label{sec:direct-effects}

Education years account for the majority of the Ed PGI's effect on labour market earnings. At correlational returns to education of around 6 percent, roughly 65 to 75 percent of the total genetic effect on earnings operates through the education channel, with a modest residual direct effect.
A sensitivity analysis varying the assumed return across a broad range shows that, at education returns above approximately 8--9 percent, the residual direct effect is no longer statistically distinguishable from zero.
These values are well within the mass of UK quasi-experimental estimates, and at these values the direct genetic effect becomes indistinguishable from zero, with the education channel accounting for the majority of Ed PGI's effect on later-life labour market outcomes.

\begin{table}[!h]
    \small
    \singlespacing
    \centering
    \caption{Direct and Indirect Effect Estimates, using Correlational Education Returns.}
    \centerline{
    \begin{tabular}{l c c c c c c c c c c c c}
        \\[-1.8ex]\hline \hline \\[-1.8ex] 
        Outcome: &
        \multicolumn{2}{c}{Occupation} & \multicolumn{2}{c}{Occupation}     & \multicolumn{2}{c}{Household income} \\ 
        & \multicolumn{2}{c}{hourly wage} & \multicolumn{2}{c}{annual income} & \multicolumn{2}{c}{(midpoint imputed)} \\
        \cmidrule(lr){2-3} \cmidrule(lr){4-5} \cmidrule(lr){6-7}
        & (1) & (2) & (3) & (4) & (5) & (6) \\
        \\[-1.8ex]\hline \\[-1.8ex]
 Education effect, $\pi$ & 0.565 & 0.551 & 0.572 & 0.557 & 0.569 & 0.559 \\ 
   & (0.037) & (0.037) & (0.040) & (0.040) & (0.039) & (0.039) \\ 
  Total genetic effect, $\theta$ & 0.050 & 0.049 & 0.042 & 0.048 & 0.056 & 0.059 \\ 
   & (0.005) & (0.005) & (0.009) & (0.008) & (0.009) & (0.008) \\ 
  Education returns, $\beta$ & 0.058 & 0.059 & 0.064 & 0.060 & 0.071 & 0.064 \\ 
   & (0.001) & (0.001) & (0.001) & (0.001) & (0.001) & (0.001) \\ 
  \midrule Direct, Ed PGI effect $\theta - \pi \beta$ & 0.018 & 0.017 & 0.006 & 0.015 & 0.015 & 0.023 \\ 
   & (0.004) & (0.004) & (0.009) & (0.008) & (0.009) & (0.008) \\ 
  Indirect, education effect $\pi \beta$ & 0.033 & 0.033 & 0.036 & 0.033 & 0.041 & 0.036 \\ 
   & (0.002) & (0.002) & (0.003) & (0.003) & (0.003) & (0.003) \\ 
  Proportion mediated through education, $\frac{\pi \beta}{\theta}$ & 0.651 & 0.653 & 0.853 & 0.693 & 0.725 & 0.611 \\ 
   & (0.056) & (0.059) & (0.224) & (0.129) & (0.130) & (0.091) \\ 
  \midrule Controls included? & No & Yes & No & Yes & No & Yes \\ 
  Observation count & 24,743 & 24,702 & 20,592 & 20,560 & 22,215 & 22,181 \\ 
  
        \\[-1.8ex]\hline \\[-1.8ex]
    \end{tabular}}
    \vspace{-0.25cm}
    \label{tab:mediate-correlational}
    \justify
    \footnotesize
    \textbf{Note:}
    This table shows the mediation point estimates, of the effect of the Ed PGI going through education years (indirect effect) and a direct genetic effect.
    Specification controls for parental Ed PGI, as mentioned in the previous section.
    All outcomes are in a log specification, and each column varies inclusion of the control variables as indicated, and standard errors on the mediation estimates are calculated by the bootstrap with 10,000 bootstrap samples.
    Observation counts vary because each outcome has a different number of observations which are non-missing.
\end{table}

Education mediates the majority of the Ed PGI's effect on earnings across all three income measures (\autoref{tab:mediate-correlational}). Using correlational returns to education, the indirect channel accounts for around 65 percent of the total genetic effect on hourly wages (column 2) and 69 percent on annual income (column 4), with a slightly wider range for household income (column 6). The residual direct genetic effect is small across all outcomes, around a third the size of the indirect effect, and the education channel dominates in each preferred specification.
The no-controls annual income estimate is a noticeable outlier at 85 percent mediated, though its wide standard error warrants treating the controlled specifications as the preferred estimates.

\begin{figure}[!h]
    \centering
    \singlespacing
    \caption{Sensitivity Analysis, Effect of Ed PGI on Log Occupational Wages via Education Years.}
    \includegraphics[width=0.85\textwidth]{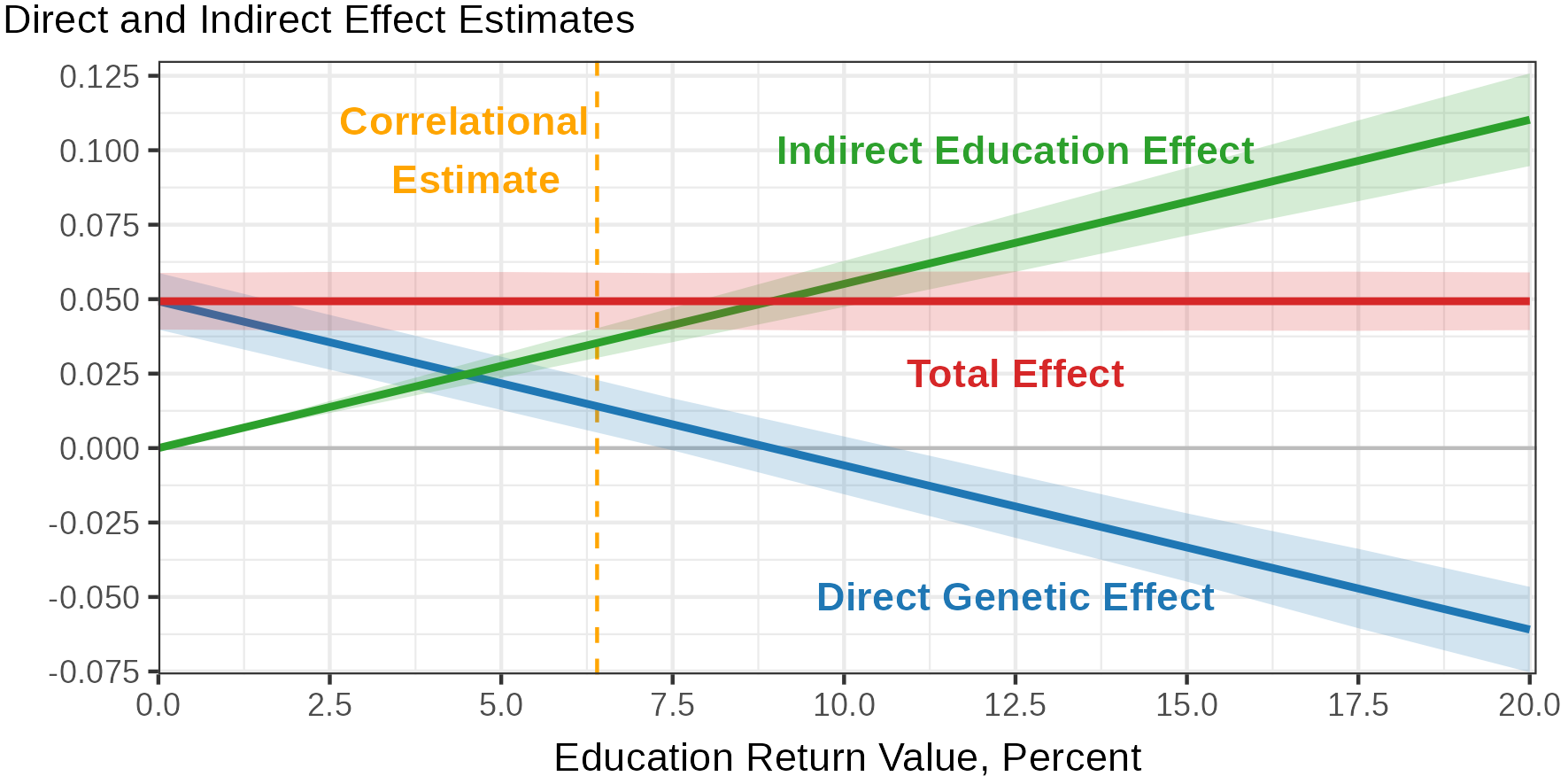}
    \label{fig:sens-hourly-wage}
    \justify
    \footnotesize
    \textbf{Note}:
    This figure shows how the indirect (education) and direct effect estimates vary using different values for average returns to education.
    The total effect refers to the average effect of the Ed PGI on log occupational hourly wages after controlling for parental Ed PGI values.
    The shaded regions refer to 95\% confidence intervals, calculated with bootstrapped standard errors clustered at the sibling family unit.
    The dotted line labelled ``Correlational Estimate'' shows the position of the correlational returns to education estimate, presented in \autoref{tab:ukb-ed-returns}.
\end{figure}

The conclusion that education years dominate is robust across the full range of credible returns to education.
\autoref{fig:sens-hourly-wage} traces the decomposition for log occupational hourly wages as the assumed return varies from zero to 20 percent.\footnote{
    The average total effect is the same as that presented in \autoref{sec:genetic-effects}, and is not varied as part of this sensitivity analysis.
}
At education returns above roughly 8 to 9 percent, the direct genetic effect becomes statistically indistinguishable from zero, and education years account for the entirety of the total effect. 
Above 10 percent the direct effect point estimate turns negative, which is difficult to reconcile with any plausible mechanism; average returns for the complier population induced by the Ed PGI are unlikely to reach that level.
The lower bound is zero by construction, because if extra education years carry no labour market return, none of the genetic effect can operate through it.
If average  returns to education were negative, then the Ed PGI channel through education would also be negative --- which is unlikely.

\begin{figure}[!h]
    \centering
    \singlespacing
    \caption{Sensitivity Analysis, Proportion of Ed PGI Effect on Income Outcomes via Education Years.}
    \centerline{
        \includegraphics[width=1.25\textwidth]{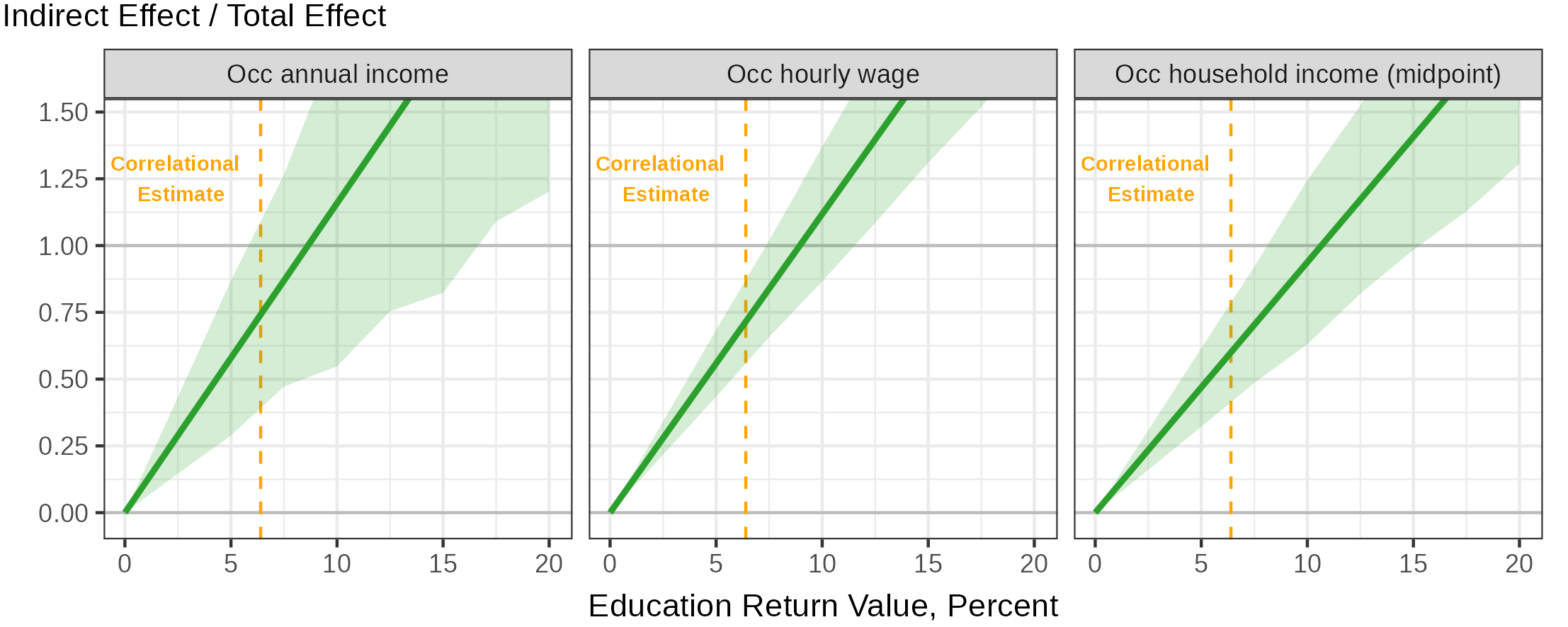}}
    \label{fig:sens-percent}
    \justify
    \footnotesize
    \textbf{Note}:
    This figure shows how the indirect (education) effect divided by total genetic effect, for each income measurement, showing percent of the Ed PGI effect mediated through education years.
    ``Occ'' refers to the word ``occupational,'' due to space constraints.
    The shaded regions refer to 95\% confidence intervals, calculated with bootstrapped standard errors.
    The specifications control for parental Ed PGI values.
    The dotted line labelled ``Correlational Estimate'' shows the position of the correlational returns to education estimate, presented in \autoref{tab:ukb-ed-returns}.
\end{figure}

The same pattern holds across all three income measures.
\autoref{fig:sens-percent} plots the ratio of the indirect effect to the total genetic effect for each outcome as the assumed return varies, and the picture is consistent: for occupational hourly wages, annual income, and household income alike, education years account for the full total genetic effect at returns of around 8 to 9 percent. The three outcomes track closely together across the range, reinforcing that the dominance of the education years channel is not specific to one of the available measures of later-life labour market income in the UKB.

\begin{table}[!h]
    \small
    \singlespacing
    \centering
    \caption{Direct and Indirect Effect Estimates, using Empirical Distribution of Education Returns Estimates.}
    \centerline{
    \begin{tabular}{l c c c c c c c c c c c c}
        \\[-1.8ex]\hline \hline \\[-1.8ex] 
        Outcome:
        & Occupation  & Occupation    & Household income \\ 
        & hourly wage & annual income & (midpoint imputed) \\
        \cmidrule(lr){2-4}
        & (1) & (2) & (3) \\
        \\[-1.8ex]\hline \\[-1.8ex]
\textbf{Education and Total Effect Estimates:} \\
 Total genetic effect, $\theta$ & 0.049 & 0.048 & 0.059 \\ 
   & (0.005) & (0.008) & (0.008) \\ 
  Education returns, $\beta$ & 0.076 & 0.076 & 0.076 \\ 
   & (0.036) & (0.036) & (0.036) \\ 
  \midrule \textbf{Mediation Estimates:} \\ 
  Direct, Ed PGI effect $\theta - \pi \beta$ & 0.008 & 0.006 & 0.017 \\ 
   & (0.021) & (0.021) & (0.021) \\ 
  Indirect, education effect $\pi \beta$ & 0.042 & 0.042 & 0.042 \\ 
   & (0.020) & (0.020) & (0.020) \\ 
  Proportion mediated through education $\frac{\pi \beta}{\theta}$ & 0.847 & 0.889 & 0.719 \\ 
   & (0.414) & (0.459) & (0.361) \\ 
  \midrule Observation count & 24,743 & 20,592 & 22,215 \\ 
  
        \\[-1.8ex]\hline \\[-1.8ex]
    \end{tabular}}
    \label{tab:mediate-boot}
    \justify
    \footnotesize
    \textbf{Note:}
    This table shows the mediation point estimates, of the effect of the Ed PGI going through education years (indirect effect) and a direct genetic effect.
    Data are resampled at the family level from the UKB analysis sample, while education returns are drawn from the distribution of causal estimates provided by \cite{patrinos2025causal} and shown in Appendix~\autoref{tab:uk-meta}.
    The regression specifications control for parental Ed PGI, and draws estimates for education returns from the distribution of causal estimates described in the previous section.
    All outcomes are in a log specification, and standard error estimates are calculated by the bootstrap with 1,000 bootstrap samples.
    Observation counts vary because each outcome has a different number of observations which are non-missing.
\end{table}

Sampling directly from the empirical distribution of UK quasi-experimental returns to education shifts the central mediation share further toward the education years channel (\autoref{tab:mediate-boot}).
Although six of the 17 IV estimates fall below the OLS anchor of around 6 percent, the distribution is right-skewed by several larger estimates, yielding an empirical average return of 7.6 percent.
At this return value, the estimated residual direct effect is statistically indistinguishable from zero across all three income measures, while education years account for approximately 72 to 89 percent of the total genetic effect.
The confidence intervals on the mediation shares are wide, but this uncertainty is inherited almost entirely from the dispersion of return estimates in the literature rather than from sampling noise in the genetic education effect.
The distribution places substantial mass around and above 7 percent, although the estimates are dispersed.

The Ed PGI's effect on labour market earnings operates primarily through education years. At correlational returns to education of around 6 percent, roughly 65 to 75 percent of the total genetic effect is mediated through the education years channel, with a residual direct effect of 35 to 25 percent.
Under the majority of plausible returns to education from the UK economics literature, this share rises further, and the direct genetic effect becomes undetectable in UKB data.

\subsection{Discussion}
\label{sec:discussion}
These results partially replicate and extend \cite{carvalho2024genetics}'s findings on the Ed PGI and labour market outcomes, and provide an account of the primary mechanism. The total genetic effects on education and earnings are broadly consistent with that paper's estimates; the mediation analysis goes further by showing that education decisions, which genetic inheritance shapes, are the primary route through which the Ed PGI translates into earnings inequality.
Individuals with a higher Ed PGI attend more years of education, and it is these additional education years, valued by the labour market, that account for the majority of their earnings advantage.
The direct genetic effect is, at most, a secondary force.

The presented mediation analysis using years of education is, if anything, a conservative account of the full role of education.
The UKB records no measures of education attainment quality, such as examination scores, ranking of university attended, or field of study; the mediator of education years completed captures only the education quantity margin.
If the Ed PGI also raises education attainment quality, then the education channel (combining quality and quantity) would be larger still, and the residual direct genetic effect correspondingly smaller.

A second limitation is that the analysis is restricted to UKB participants with at least one sibling in the data.
This sample cannot be generalised to the broader British population, and the sibling-imputation procedure introduces measurement error relative to the ideal case of observing both parental Ed PGI values directly.
A subtler issue is that when parents are far apart in the Ed PGI distribution, the range of values a child can inherit is wider; this additional variation in expected inheritance is not fully captured by the sibling-imputed procedure.

Future research would benefit from data that observe both parents' Ed PGI values directly, removing the measurement error introduced by sibling imputation and improving the precision of the genetic education effect.
Progress on the quality margin would require data measuring any of the education quality channels.
Together, these extensions would sharpen both the causal estimates and the mediation analysis presented here.

\section{Summary and Concluding Remarks}
\label{sec:conclusion}
This paper uses quasi-random variation in genetic inheritance to estimate the causal effects of the Ed PGI on education and labour market outcomes, and decomposes those effects into a direct genetic channel and an indirect channel operating through education years in a causal mediation framework.
The Ed PGI raises both years of education and later-life income by multiple measures, replicating the core findings of the recent literature \citep{carvalho2024genetics}.
A causal mediation analysis uses these findings as a basis, and at plausible returns to education from the economics literature for Britain, the majority of the total Ed PGI income effect operates through the education years channel, with a residual direct genetic effect that is small and, under higher return values, indistinguishable from zero.
The literature had noted suggestive evidence that direct effects may exist, but had not delivered a quantitative decomposition; this paper does so, and finds that the education years channel accounts for the bulk of the total effect across the plausible range of returns, with (at most) small direct effects.


\singlespacing
\bibliographystyle{agsm}
\bibliography{sections/08-bibliography-doi.bib}

\newpage
\appendix
\setcounter{table}{0}
\renewcommand{\thetable}{A\arabic{table}}
\setcounter{figure}{0}
\renewcommand{\thefigure}{A\arabic{figure}}

\section{Supplementary Appendix}
\label{appendix}
This project used computational tools which are fully open-source.
Any comments or suggestions may be sent to me at \href{mailto:seh325@cornell.edu}{\nolinkurl{seh325@cornell.edu}}.

Data analysis with the \textit{Tidyverse} \citep{tidyverse} package, for the R language \citep{R2023}, made the empirical analysis for this paper possible.

\subsection{UK Biobank Dataset Construction}

\subsubsection{Identifying Family Clusters by Genetic Similarity}
\label{appendix:identifying-families}

The UKB does not directly identify participants' family members, so I define family links using the genetic relatedness data.
I follow the approach in \citet{muslimova2025environment}, classifying genetically related pairs using the kinship coefficient and the identity-by-state zero coefficient (IBS$_0$).
These thresholds come from the KING relationship-inference procedure used to construct the UKB kinship matrix, which identifies genetically related pairs up to third-degree relatives \citep{manichaikul2010robust}.
The kinship coefficient measures the degree of genetic relatedness between two individuals, while IBS$_0$ measures the fraction of genotyped markers for which the two individuals share no alleles.
These two measures jointly distinguish different forms of close genetic relatedness.

\begin{table}[!h]
    \singlespacing
    \centering
    \small
    \caption{Genetically Identified Family Links in the UKB.}
    \begin{tabular}{l c c c}
        \\[-1.8ex]\hline \hline \\[-1.8ex] 
        Genetic link & Kinship Coefficient & IBS$_0$ & Count \\
        \\[-1.8ex]\hline \hline \\[-1.8ex] 
 Duplicate / Monozygotic twins & $> 0.3540$ &  & 179 \\ 
  1st degree / Parent-child & $0.1770$--$0.3540$ & $< 0.0012$ & 6,249 \\ 
  1st degree siblings & $0.1770$--$0.3540$ & $\geq 0.0012$ & 22,614 \\ 
  2nd--3rd degree relatives / cousins & $0.0442$--$0.1770$ &  & 77,878 \\ 
  \hline Total &  &  & 106,920 \\ 
  
        \\[-1.8ex]\hline \\[-1.8ex]
    \end{tabular}
    \vspace{-0.25cm}
    \label{tab:ukb-relations-count}
\end{table}

I classify duplicate observations and monozygotic twins as pairs with kinship coefficient above 0.3540.
First-degree relatives are pairs with kinship coefficient between 0.1770 and 0.3540.
Within this first-degree interval, parent-child pairs and full siblings are distinguished by IBS$_0$.
Parent-child pairs have IBS$_0$ below 0.0012, since a biological parent and child should share one allele at nearly every locus.
Full siblings have IBS$_0$ weakly above 0.0012, since full siblings may share zero, one, or two alleles at a given locus.
Second- and third-degree relatives are defined as pairs with kinship coefficient between 0.0442 and 0.1770.

\autoref{tab:ukb-relations-count} reports the number of genetically related pairs in each category, among all pairs of individuals with kinship coefficient at least 0.0442, corresponding to cousins or closer relatives.
The analysis sample includes individuals with at least one genetically identified first-degree connection in the UKB relatedness file, and filtered to those with non-missing data in occupation codes and education variables.
The counts differ from \citet{muslimova2025environment} thanks to this filtering.

I use these pairwise genetic links to construct family identifiers. Each UKB participant is treated as a node in a family-relatedness network, and each genetically identified relationship is treated as an edge between two participants. Family identifiers are then assigned from the connected components of this network. The primary analysis sample is based on individuals with at least one genetically identified full sibling in the UKB, since full-sibling links are the relevant family relationship for the within-family design and the sibling-based parental PGI imputation procedure.

The design of the UKB allows for a larger number of family clusters than would be expected from a random sample of the British population.
This is because study enrolment letters were sent to households, and the household head (or to whomever the letter was addressed) could optionally enrol their family members in addition to themselves.

\subsubsection{Occupation Wage Imputation}
\label{appendix:pgi-impute}

The UKB does not report continuous individual wages, but it does record labour-market information that can be used to impute wages, including 4-digit Standard Occupational Classification (SOC) codes and reported working hours.
I construct occupation-based wages using the supplementary files and replication code from \cite{kweon2025associations}, following the same imputation procedure as \citet{carvalho2024genetics}.
The resulting measure should be interpreted as an occupation-based prediction of labour-market earnings, rather than a directly observed individual wage.

The imputation procedure combines information from three UK data sources.
First, the Labour Force Survey (LFS) provides individual-level wage data and labour-market covariates.
Second, the Annual Survey of Hours and Earnings (ASHE) provides high-quality occupation-level wage information by SOC code.
Third, the British Household Panel Survey (BHPS) is used to validate the imputation procedure in an independent dataset.
The procedure estimates a parametric prediction model for log wages in the LFS, using ASHE mean and median wages for each occupation group together with demographic and labour-market predictors.
The estimated model is then applied to UKB participants using their SOC code and observed labour-market characteristics.

The key predictor is the participant's 4-digit SOC code.
Rather than including a full set of 4-digit occupation fixed effects, which would create sparse cells after interacting occupation with age, sex, full-time status, and year, the \cite{kweon2025associations} procedure uses ASHE mean and median wages for each occupation group as continuous predictors.
This preserves detailed occupational wage information while avoiding failed or noisy imputations in small occupation-demographic cells.
I use the imputed hourly wage measure, and construct annual occupation income by multiplying the imputed hourly wage by reported weekly hours and 52 weeks.

\subsection{Genetic Correlations}

\begin{figure}[H]
    \centering
    \singlespacing
    \caption{Correlation for Ed PGI, and Ed PGI Minus Parental Mean, with Demographic Variables.}
    \label{fig:demographic-correlates}
    \includegraphics[width=0.85\textwidth]{
        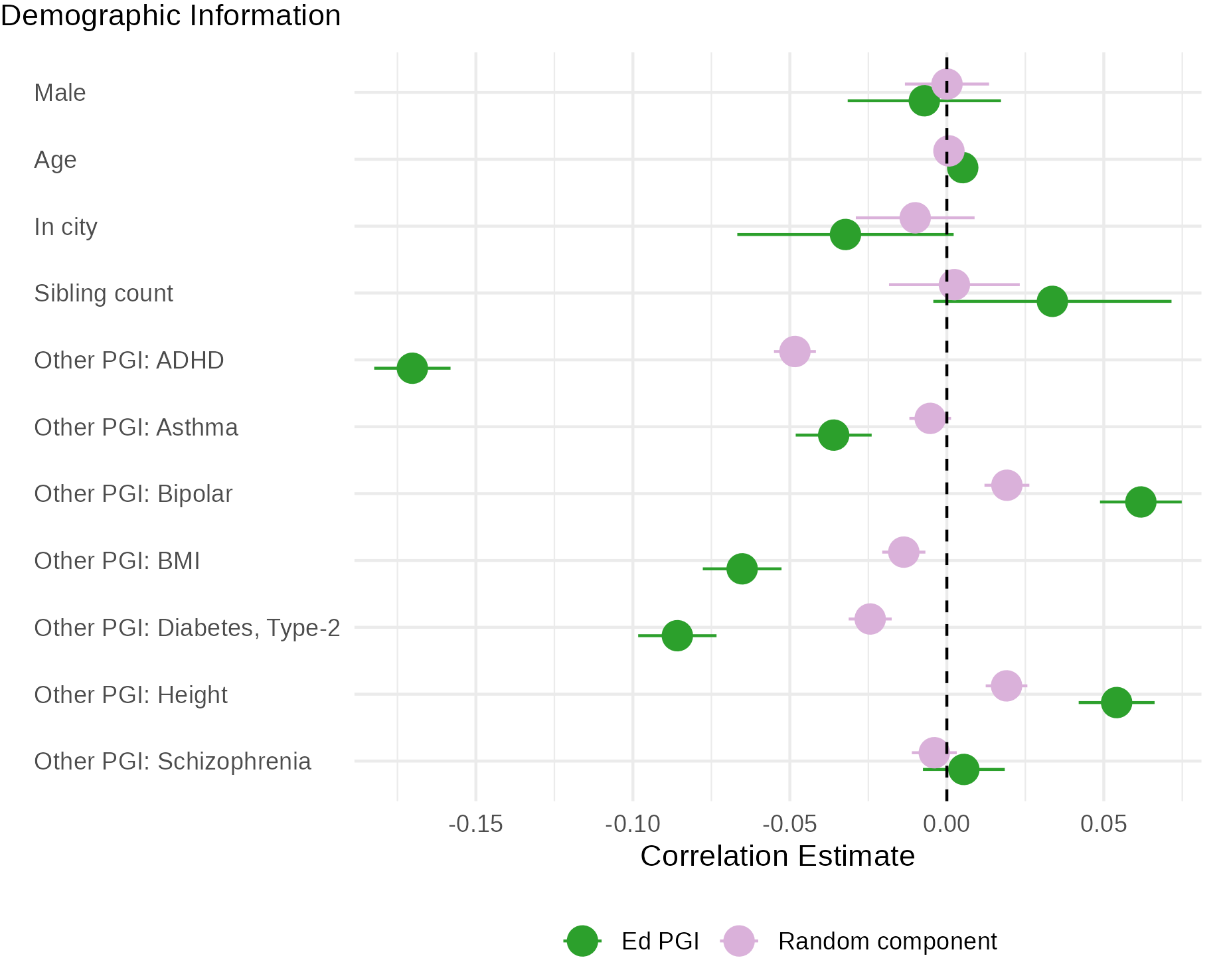}
    \justify
    \footnotesize
    \textbf{Note}:
    Multivariate regression estimates for the linear model $Z_i = \vec \phi\,' \, \vec X_i + \eta_i$, where $\vec X_i$ is the vector demographic variables described by the labels on the $y$-axis.
    The dots are the point estimates, and bars are the 95\% confidence intervals.
    The random component refers to the Ed PGI after residualing out parental values,
\end{figure}

\subsection{Parametric Education--Mediation Identification}
\label{appendix:mediation-identification}
This appendix expands on the parametric identification argument used in \autoref{sec:direct}.
The main text writes the indirect effect as the product of the average education effect and the returns to education.
This representation is convenient, but it imposes structure: the unrestricted chain-rule object is an average of individual-level products, not mechanically the product of two averages.

Recall the indirect effect,
\[
    \text{Indirect Effect}
    =
    \E{
        \partialdiff{z}
        Y_i(Z_i^\text{\textcolor{childColour}{Child}}, D_i(z))
    }.
\]
Holding the direct value of the Ed PGI fixed at the realised child value, the individual-level chain rule gives
\[
    \partialdiff{z}
    Y_i(Z_i^\text{\textcolor{childColour}{Child}}, D_i(z))
    =
    \left.
    \partialdiff{d}
    Y_i(Z_i^\text{\textcolor{childColour}{Child}}, d)
    \right|_{d = D_i(z)}
    \times
    \partialdiff{z}D_i(z).
\]
Therefore, the average indirect effect is
\begin{equation}
    \label{eqn:appendix-indirect-product}
    \text{Indirect Effect}
    =
    \E{
        \left.
        \partialdiff{d}
        Y_i(Z_i^\text{\textcolor{childColour}{Child}}, d)
        \right|_{d = D_i(z)}
        \times
        \partialdiff{z}D_i(z)
    }.
\end{equation}
This is the most general parametric chain-rule object.
It shows that the relevant return to education is not necessarily the population-average return to schooling.
Instead, the relevant return is weighted by the individuals and education margins whose schooling is shifted by the Ed PGI.

Define the average education effect as
\[
    \pi
    =
    \E{\partialdiff{z}D_i(z)}.
\]
Provided that $\pi \neq 0$, \autoref{eqn:appendix-indirect-product} can be written as
\begin{equation}
    \label{eqn:appendix-beta-mediated}
    \text{Indirect Effect}
    =
    \pi \beta^M,
\end{equation}
where
\begin{equation}
    \label{eqn:appendix-beta-mediated-definition}
    \beta^M
    =
    \frac{
        \E{
            \left.
            \partialdiff{d}
            Y_i(Z_i^\text{\textcolor{childColour}{Child}}, d)
            \right|_{d = D_i(z)}
            \times
            \partialdiff{z}D_i(z)
        }
    }{
        \E{\partialdiff{z}D_i(z)}
    }.
\end{equation}
The parameter $\beta^M$ is the mediator-relevant return to education: the return to an additional year of education averaged over the education margins induced by the Ed PGI, with weights proportional to each individual's education response $\partial D_i(z)/\partial z$.
If the Ed PGI shifts education more strongly for individuals with high returns to schooling, then $\beta^M$ exceeds the unweighted average return.
If it shifts education more strongly for individuals with low returns, then $\beta^M$ is lower.

The weights in this average are proportional to each individual's education response,
$\partial D_i(z)/\partial z$.
If a higher Ed PGI weakly increases education for every individual,
\begin{equation}
    \label{eqn:appendix-education-monotonicity}
    \partialdiff{z}D_i(z) \geq 0
    \qquad \text{for all } i,
\end{equation}
then $\beta^M$ is a positively weighted average of individual returns to education.
This monotonicity condition is not required for the chain-rule identity in
\autoref{eqn:appendix-indirect-product}.
Without it, the decomposition remains valid, but individuals with
$\partial D_i(z)/\partial z < 0$ receive negative weight in $\beta^M$, so that
$\beta^M$ need not lie within the support of individual education returns
\citep{blackwell2024assumption}.

The main-text expression,
\[
    \text{Indirect Effect}
    =
    \pi \beta,
\]
therefore requires an additional separability condition linking the mediator-relevant return $\beta^M$ to an externally supplied return-to-education parameter $\beta$.
One sufficient condition is homogeneous returns to education,
\[
    \left.
    \partialdiff{d}
    Y_i(Z_i^\text{\textcolor{childColour}{Child}}, d)
    \right|_{d = D_i(z)}
    =
    \beta
    \qquad \text{for all } i.
\]
Under this condition,
\[
    \E{
        \left.
        \partialdiff{d}
        Y_i(Z_i^\text{\textcolor{childColour}{Child}}, d)
        \right|_{d = D_i(z)}
        \times
        \partialdiff{z}D_i(z)
    }
    =
    \beta
    \E{\partialdiff{z}D_i(z)}
    =
    \pi\beta.
\]
A weaker sufficient condition is that the education return and the Ed-PGI-induced education response have zero covariance,
\begin{equation}
    \label{eqn:appendix-separability}
    \E{
        \left.
        \partialdiff{d}
        Y_i(Z_i^\text{\textcolor{childColour}{Child}}, d)
        \right|_{d = D_i(z)}
        \times
        \partialdiff{z}D_i(z)
    }
    =
    \E{
        \left.
        \partialdiff{d}
        Y_i(Z_i^\text{\textcolor{childColour}{Child}}, d)
        \right|_{d = D_i(z)}
    }
    \times
    \E{\partialdiff{z}D_i(z)}.
\end{equation}
This condition permits heterogeneous returns and heterogeneous education responses, but requires their individual-level covariance to equal zero.
Under \autoref{eqn:appendix-separability}, the indirect effect is again
\[
    \text{Indirect Effect}
    =
    \pi\beta,
    \qquad
    \beta
    =
    \E{
        \left.
        \partialdiff{d}
        Y_i(Z_i^\text{\textcolor{childColour}{Child}}, d)
        \right|_{d = D_i(z)}
    }.
\]

The interpretation of $\beta$ is therefore central.
The Mendelian design identifies $\pi$, the average causal effect of the Ed PGI on education years.
It also identifies $\theta$, the average total effect of the Ed PGI on labour-market outcomes.
It does not, by itself, identify the return to education relevant for the education margins shifted by the Ed PGI.
The mediation decomposition therefore treats $\beta$ as the remaining input.
The empirical analysis first uses the correlational return in the UKB as a transparent benchmark, and then varies $\beta$ over values from quasi-experimental estimates in the returns-to-education literature.

Given a value of $\beta$, the indirect and direct effects are
\[
    \text{Indirect Effect}
    =
    \pi\beta,
    \qquad
    \text{Direct Effect}
    =
    \theta - \pi\beta.
\]
Thus, the causal content of the mediation analysis rests on two pieces.
First, $\pi$ and $\theta$ are identified by quasi-random Mendelian variation in the child Ed PGI, conditional on parental Ed PGI.
Second, the chosen value of $\beta$ must be interpreted as the return to education relevant for the education years induced by the Ed PGI.
The sensitivity analysis makes this second requirement explicit by showing how the direct and indirect effects change as $\beta$ varies across plausible causal returns to education.

\subsection{Returns to Education by the MSLA Rise}
\label{appendix:MSLA}
The UK Biobank includes a total of 500,000 members of the British public, with a roughly even distribution of birth years over 1945--1965.
In 1972, the British government increased the Minimum School Leaving Age (MSLA) by one year, from age 15 to 16.
The change took effect at the start of the 1972-73 school year, and applied to those born on or after 1 September 1957.
The reform therefore induced extra school years for individuals born around date 1 September 1957.
British school children sit exams for the GCSE qualification at age 16, so the reform pushed a larger share of the UK Biobank cohorts to remain in school to 16 and to complete these qualifications.

The 1972 MSLA change has been previously used for labour market education returns \citep{oreopoulos2006estimating}, and (null) mortality effects of compulsory education \citep{clark2013effect}.
The UK Biobank population overlaps well with the birth date cutoff, and so has been used in a series of papers for estimating education returns in the UK Biobank, for health outcomes \citep{barcellos2023distributional,barcellos2025education,davies2018causal} and occupation imputed income \citep{barcellos2021effect}.

\begin{figure}[H]
    \centering
    \singlespacing
    \caption{Mean Effects of the MSLA Rise.}
    \begin{subfigure}[b]{0.495\textwidth}
        \centering
        \includegraphics[width=\textwidth]{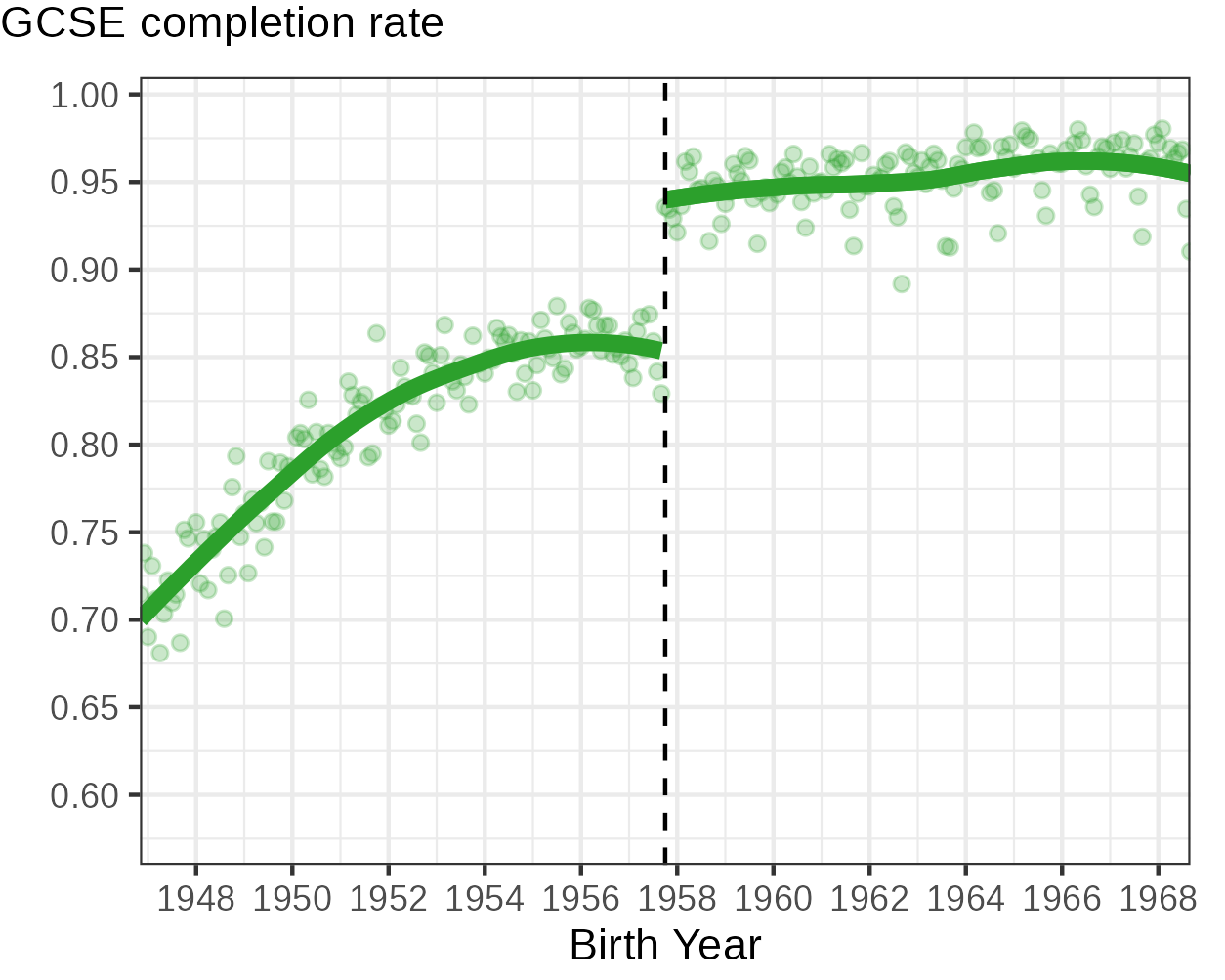}
    \end{subfigure}
    \begin{subfigure}[b]{0.495\textwidth}
        \centering
        \includegraphics[width=\textwidth]{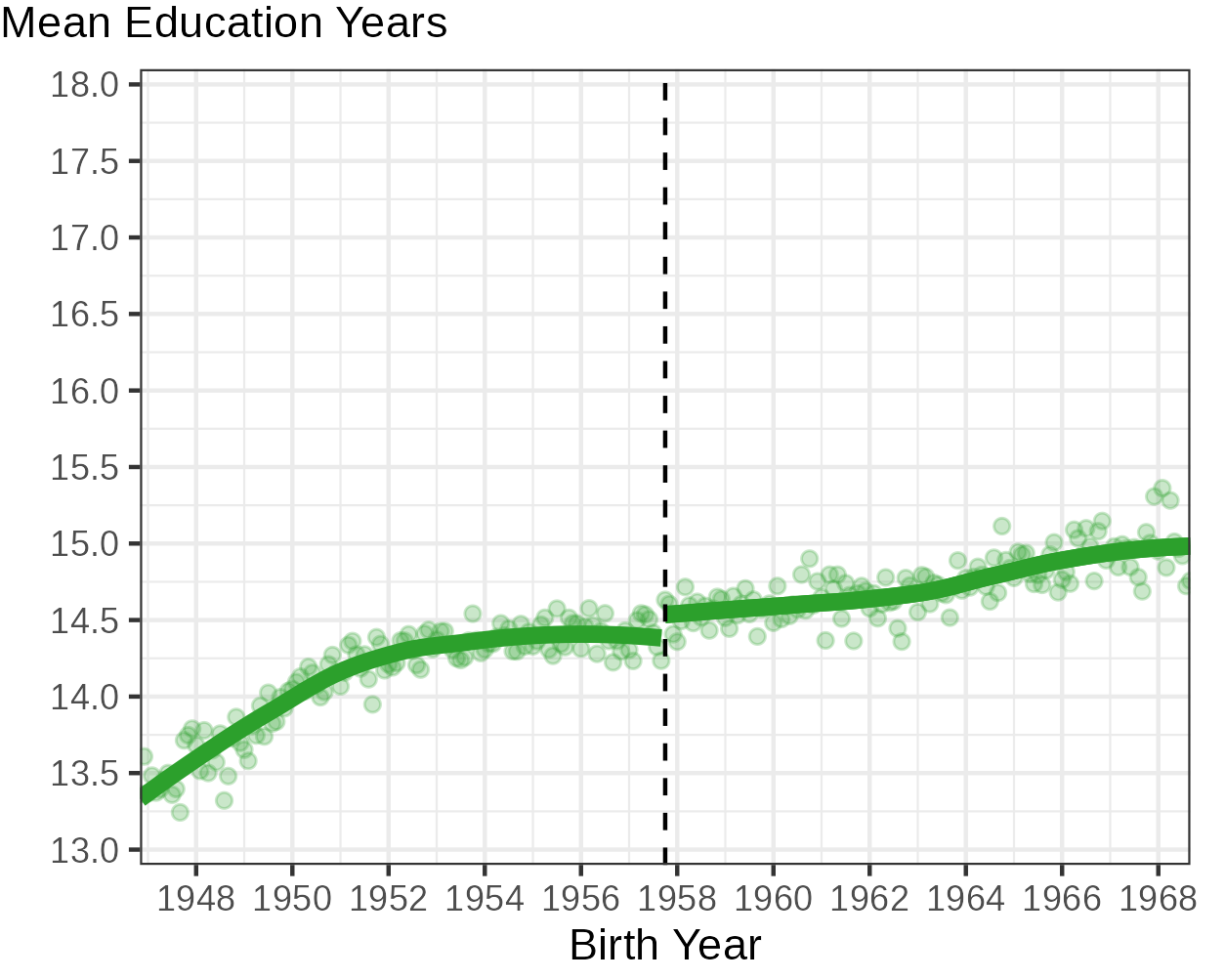}
    \end{subfigure}
    \begin{subfigure}[b]{0.495\textwidth}
        \centering
        \includegraphics[width=\textwidth]{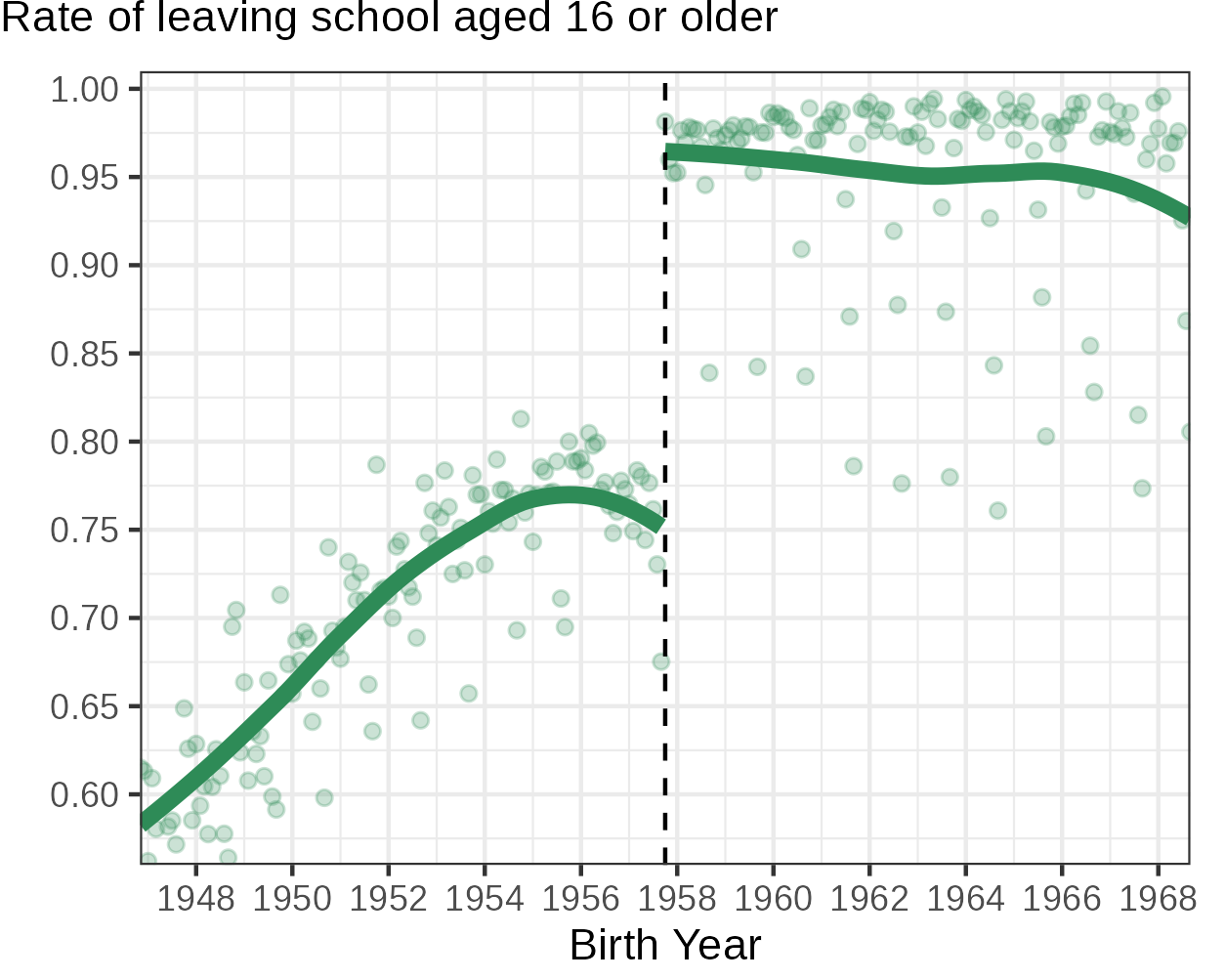}
    \end{subfigure}
    \begin{subfigure}[b]{0.495\textwidth}
        \centering
        \includegraphics[width=\textwidth]{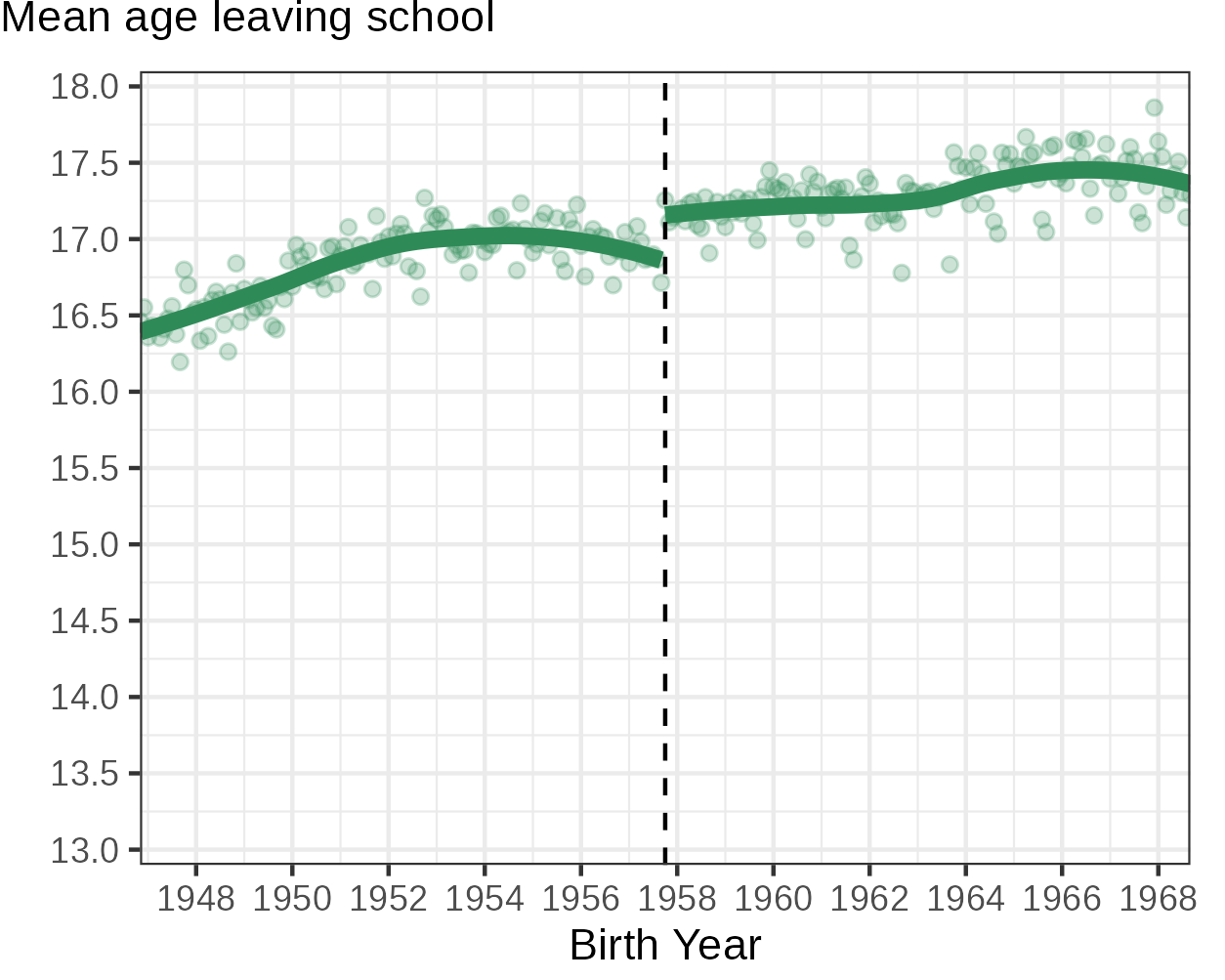}
    \end{subfigure}
    \begin{subfigure}[b]{0.495\textwidth}
        \centering
        \includegraphics[width=\textwidth]{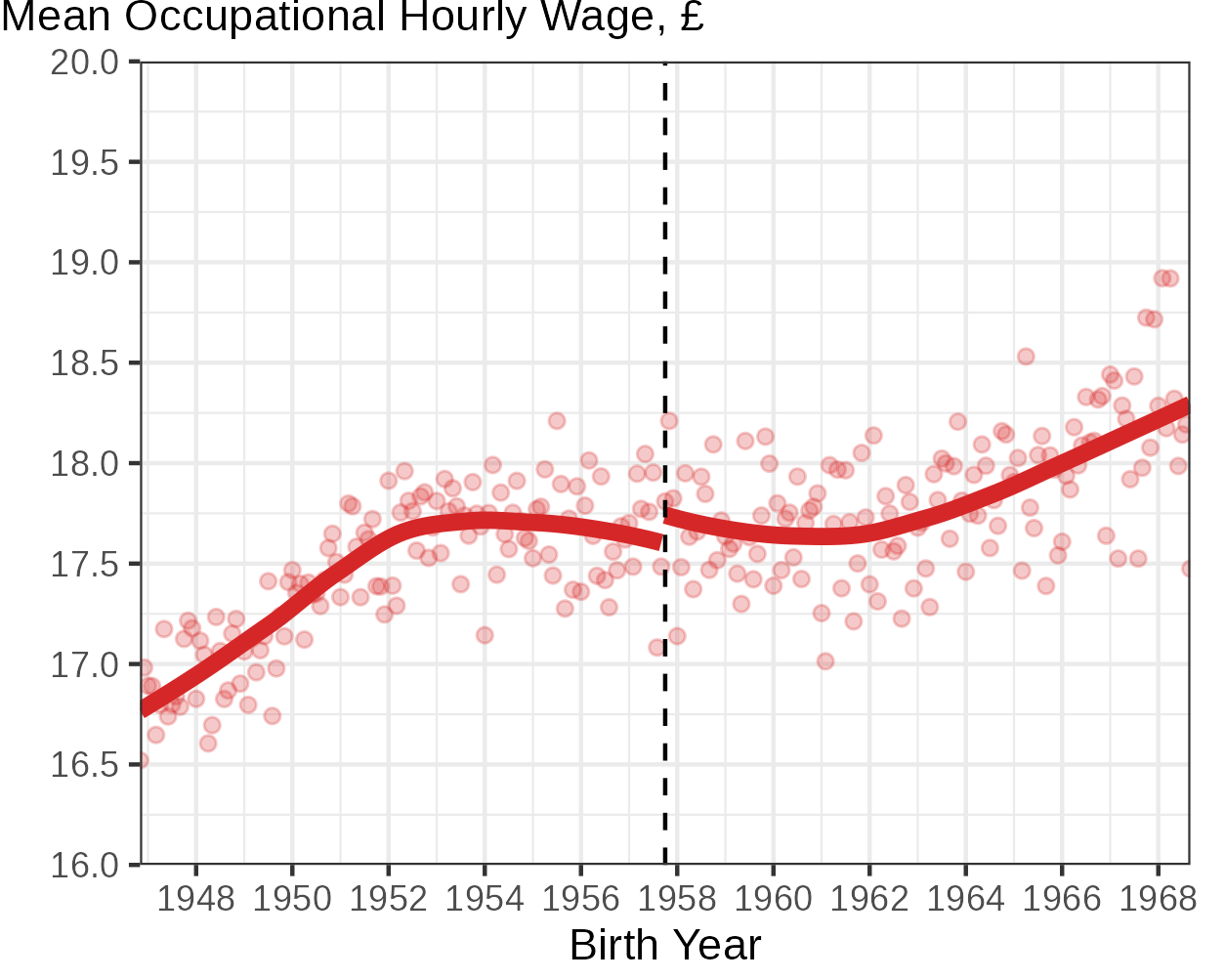}
    \end{subfigure}
    \begin{subfigure}[b]{0.495\textwidth}
        \centering
        \includegraphics[width=\textwidth]{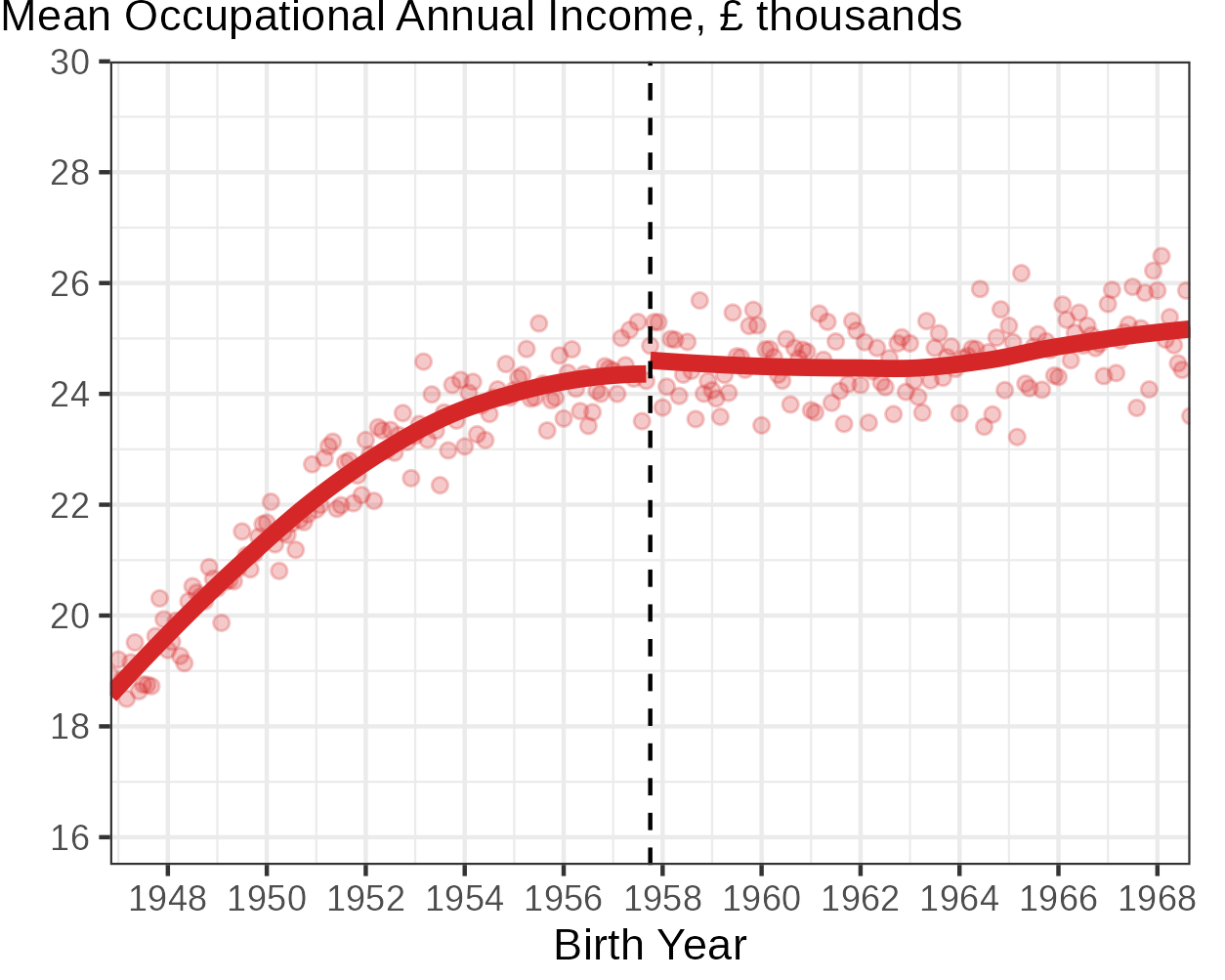}
    \end{subfigure}
    \label{fig:mlsa-nl}
    \vspace{-0.5cm}
    \justify
    \footnotesize
    \textbf{Note}:
    These figures show the effects of the 1972 MSLA rise on multiple outcomes, where individuals born after September 1957 (the dashed line) were legally compelled to stay in school until age 16, whereas before only until age 15.
    Each dot is the mean for UK Biobank respondents with non-missing education data born in that month-year, and the lines are local polynomials either side of September 1957.
\end{figure}

Birth month is the running variable in a fuzzy Regression Discontinuity Design (RDD), using the indicator for being born September 1957 and after as an instrument.
Because not all those eligible necessarily stayed beyond 15, the post-cutoff indicator instruments education variables, where mentioned.
I estimate the following RDD estimands, estimating the effect of the MSLA change on education variables --- represented by $D_i$.
\[ \lim_{v \to^- 1957.75} \Egiven{D_i}{V_i = v}, \;\;\;\;
    \lim_{v \to^+ 1957.75} \Egiven{D_i}{V_i = v} \]
where $V_i$ represents individual $i$'s birth year plus birth month divided by 12 (so that 1957.75 represents born in September 1957), and $\to^-$ refers to the limit from below, $\to^+$ above.
The goal is to identify the average compliance rate of the MSLA rise,
\[ \E{D_i(1) - D_i(0)}
    = \lim_{v \to^+ 1957.75} \Egiven{D_i}{V_i = v}
        - \lim_{v \to^- 1957.75} \Egiven{D_i}{V_i = v}. \]

The MSLA rise significantly increases education years, completing of age 16 secondary school qualifications (GCSEs) --- these are measured of education coded by qualifications, as used by \cite{okbay2022polygenic} in calculating the Ed PGI.
It had an even larger effect on age at leaving school, which is the basis of the education variables that \cite{barcellos2021effect} focuses on.
The mean of these variables by birth month are shown in \autoref{fig:mlsa-nl}.
Note that the economics literature has argued that null estimates for education returns in the MSLA design are a result of the MSLA rise inducing a low-quality margin of education \citep{clark2013effect}, so while age leaving education variables are more significant for this instrument they may not be economically relevant.

I also use the RDD design to measure changes in later labour market outcomes.
The UK Biobank asks questions of the sector people work in, giving high detail occupation codes (i.e., SOC codes).
Using this value, and other demographic details, I impute occupation coded income for the UK Biobank respondents using data from the British Annual Survey of Hours and Earnings.\footnote{
    Thanks to \cite{kweon2025associations} for sharing code on this imputation.
}
This gives occupation coded hourly income, and annual wage calculated from hours worked as outcome measures in the MSLA rise, affected by the MSLA rise
\[ \lim_{v \to^+ 1957.75} \Egiven{Y_i}{V_i = v}
    - \lim_{v \to^- 1957.75} \Egiven{Y_i}{V_i = v}. \]
The final panels of \autoref{fig:mlsa-nl} show these outcomes, and this reduced-form effect of the MSLA is not significantly different from zero.
\cite{barcellos2021effect} analyse this same RDD with UKB data.

\begin{table}[H]
    \singlespacing
    \small
    \centering
    \caption{Estimates of the MSLA Change Effects.}
    \centerline{
    \begin{tabular}{l c c c c c c c c}
        \\[-1.8ex]\hline \hline \\[-1.8ex]
        \textbf{Panel A: First-stage.}
        & \multicolumn{4}{c}{First-stage.}
            & \multicolumn{4}{c}{Reduced form.}\\
        \cmidrule(lr){2-5} \cmidrule(lr){6-9}
        Outcome:
        & \multicolumn{2}{c}{GCSE}       & \multicolumn{2}{c}{Education} & \multicolumn{2}{c}{Hourly} & \multicolumn{2}{c}{Annual} \\
        & \multicolumn{2}{c}{completion} & \multicolumn{2}{c}{years}     & \multicolumn{2}{c}{wage}   & \multicolumn{2}{c}{income} \\
        \cmidrule(lr){2-5} \cmidrule(lr){6-9}
        & (1) & (2) & (3) & (4) & (5) & (6) & (7) & (8) \\
        \\[-1.8ex]\hline \\[-1.8ex]
 MSLA Effect & 0.089 & 0.247 & 0.158 & 0.314 & 0.006 & 0.006 & 0.010 & 0.010 \\ 
    & (0.006) & (0.011) & (0.060) & (0.058) & (0.007) & (0.007) & (0.010) & (0.010) \\ 
  \\ $F$-statistics & 216 & 530 & 6.85 & 29.7 & 0.939 & 0.939 & 0.830 & 0.830 \\ 
  CCT bandwidth & 2.66 & 1.60 & 3.21 & 2.60 & 4.10 & 4.10 & 3.87 & 3.87 \\ 
  Bandwidth observations & 64,011 & 24,185 & 77,881 & 38,718 & 99,646 & 99,646 & 93,885 & 93,885 \\ 
  
        \\[-1.8ex]\hline \\[-1.8ex]
        \textbf{Panel B: Second-stage.}
            & \multicolumn{4}{c}{OLS.}
                & \multicolumn{4}{c}{MSLA IV.} \\
        \cmidrule(lr){2-5} \cmidrule(lr){6-9}
        & (1) & (2) & (3) & (4) & (5) & (6) & (7) & (8) \\
        \\[-1.8ex]\hline \\[-1.8ex]
        \multicolumn{8}{l}{Outcome: Log Occupation Hourly Wage} \\
 GCSE completion & 0.372 & 0.193 &   &   & 0.070 & 0.068 &   &   \\ 
    & (0.008) & (0.008) &   &   & (0.094) & (0.039) &   &   \\ 
  Education years &   &   & 0.061 & 0.034 &   &   & 0.044 & 0.057 \\ 
    &   &   & (0.000) & (0.002) &   &   & (0.047) & (0.029) \\ 
  \\ CCT bandwidth & 10.0 & 10.0 & 10.0 & 10.0 & 2.54 & 2.71 & 4.21 & 4.08 \\ 
  Bandwidth observations & 245,740 & 149,732 & 245,740 & 149,732 &  62,027 &  39,938 & 101,570 &  58,987 \\ 
  
        \\[-1.8ex]\hline \\[-1.8ex]
        \multicolumn{8}{l}{Outcome: Log Occupation Annual Income} \\
 GCSE completion & 0.392 & 0.188 &   &   & 0.100 & 0.046 &   &   \\ 
    & (0.015) & (0.009) &   &   & (0.151) & (0.065) &   &   \\ 
  Education years &   &   & 0.065 & 0.032 &   &   & 0.062 & 0.048 \\ 
    &   &   & (0.001) & (0.002) &   &   & (0.080) & (0.049) \\ 
  \\ CCT bandwidth & 10.0 & 10.0 & 10.0 & 10.0 & 2.61 & 2.86 & 3.84 & 4.16 \\ 
  Bandwidth observations & 245,740 & 149,732 & 245,740 & 149,732 &  64,011 &  42,307 &  93,885 &  60,155 \\ 
  \\[-1.8ex]\hline \\[-1.8ex] Qualification definition & Yes &   & Yes &   & Yes &   & Yes &   \\ 
  School age definition &   & Yes &   & Yes &   & Yes &   & Yes \\ 
  
        \\[-1.8ex]\hline \hline \\[-1.8ex]
    \end{tabular}
    }
    \vspace{-0.25cm}
    \label{tab:mlsa-iv}
    \justify
    \footnotesize
    \textbf{Note:}
    These point estimates are calculated among all UK Biobank participants born within a bandwidth of the MSLA rise cutoff (not just the sibling subsample).
    The odd columns use a definition of education attainment (age 16 schooling) based on reported education qualifications; even columns report based on school leaving age.
    The Panel B OLS results show estimates of an OLS specification for the correlation between education and wages/income with local polynomial trends each side of 1957.75 birthdate, attempting to mimic an RD set-up with a hard-coded bandwidth of 10 years for comparison.
\end{table}

I estimate this RDD equation with triangular kernel weights (weighting closer to the cutoff), an optimal bandwidth \citep{calonico2014robust,calonico2015rdrobust}, and non-parametric controls on either side of the cutoff \citep{gelman2019high}.
These estimates are collected in \autoref{tab:mlsa-iv}.

\begin{figure}[H]
    \centering
    \singlespacing
    \caption{Fuzzy RDD Estimates, Sensitivity to Bandwidth Choice.}
    \begin{subfigure}[b]{0.495\textwidth}
        \centering
        \includegraphics[width=\textwidth]{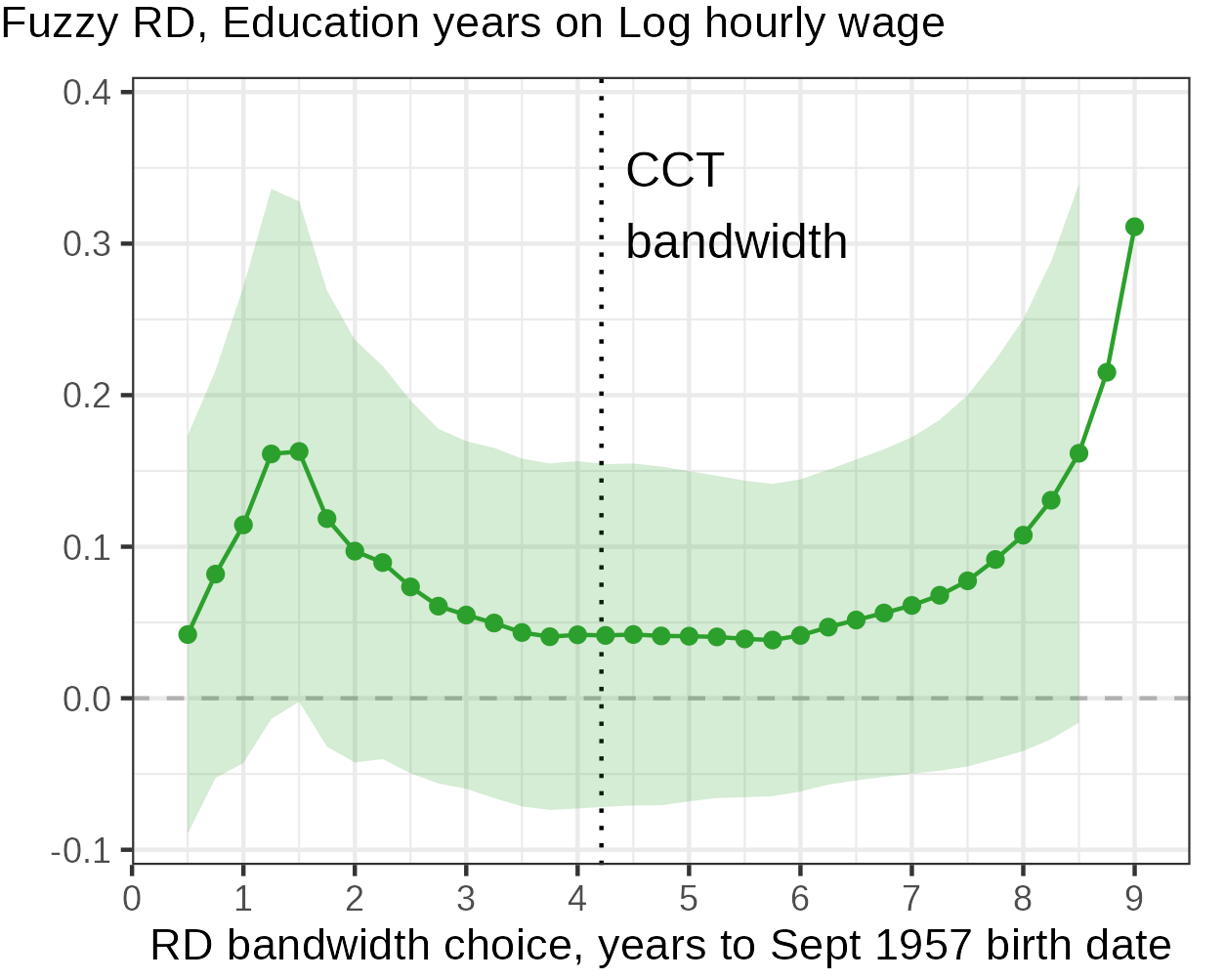}
    \end{subfigure}
    \begin{subfigure}[b]{0.495\textwidth}
        \centering
        \includegraphics[width=\textwidth]{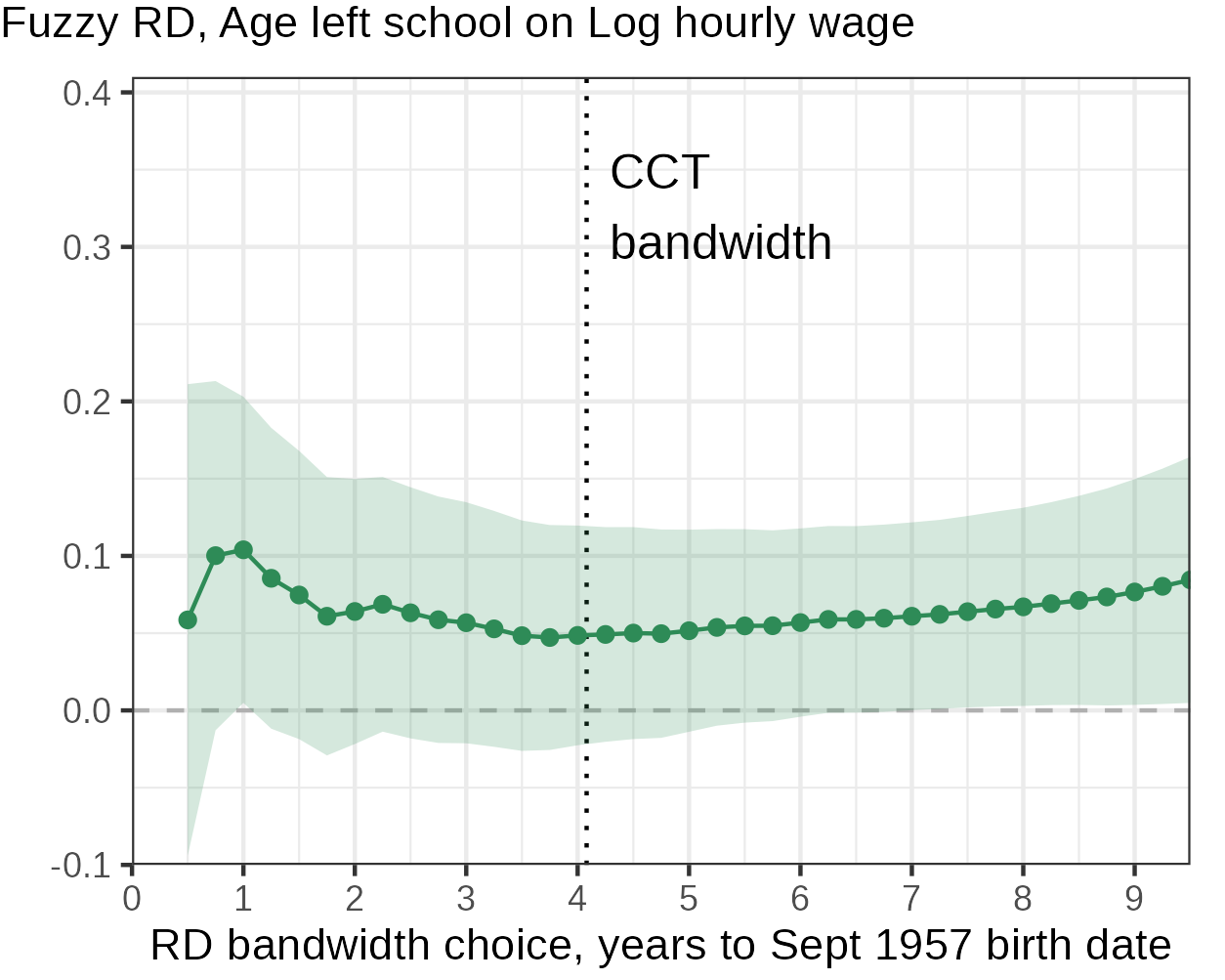}
    \end{subfigure}
    \label{fig:mlsa-bw}
    \vspace{-0.5cm}
    \justify
    \footnotesize
    \textbf{Note}:
    These figures show the sensitivity of fuzzy RD estimates using the MSLA discontinuity as an IV for education years (qualification measure) or age leaving school, using a different bandwidth choice for each.
    The shaded regions are the 95\% confidence interval.
    The dotted line with CCT refers to the \cite{calonico2014robust} optimal bandwidth choice, calculated using \textit{rdrobust} in \textit{R} \citep{calonico2015rdrobust}.
\end{figure}

The MSLA rise increased years of education, as measured in both qualification definitions and school leaving age definition.
However, the outcome of occupation coded wages and annual income are not strong; correspondingly, the IV estimates using the MSLA rise for education returns are not distinguishable from zero.
This holds true for a range of bandwidth choices, shown in \autoref{fig:mlsa-bw}.

\subsection{Instrumental Variables Connection}
\label{appendix:iv}
The CM analysis in this paper is conceptually similar to an instrumental variables analysis; the main difference is that it does not assume the exclusion restriction, and instead attempts to model and estimate this direct effect.
One might consider using the Ed PGI as an instrument for education years, interpreting the reduced-form effect of the Ed PGI on earnings only through its first-stage effect on education. 
This is the logic of Mendelian Randomisation (MR), where genetic variants are used as instruments for health conditions, and any direct effect of the genetic index on the outcome is treated as a violation of the exclusion restriction.

\cite{widding2026}, for example, uses the Ed PGI as an instrument for education in an MR design; this approach is not well suited to the present question as it assumes the direct effects are exactly zero for every individual, giving estimates for returns to education contaminated by plausibly non-zero direct effects.
The CM approach instead estimates both the direct genetic effect and the indirect education effect.

There is a literature for modelling such exclusion violations in economics \citep{conley2012plausibly,kolesar2015}, and in genetic settings using alternative assumptions tailored to MR \citep{bowden2015mendelian,spiller2019detecting}.
These alternative approaches require structure on how invalid instruments violate exclusion; the most common adjustment for genetic settings, MR-egger \citep{bowden2015mendelian}, relies on the direct genetic effects being uncorrelated with instrument strength, paralleling the invalid instruments assumption in the econometrics literature \citep{kolesar2015}. 
This is difficult to justify in the present setting, where the SNPs that most strongly predict education could have correspondingly larger direct effects on earnings through cognitive skills, non-cognitive traits, health, field of study, occupational sorting, or other labour-market relevant channels.

\subsection{Returns to Education Literature}
\begin{table}[!htbp]
    \small
    \singlespacing
    \centering
    \vskip-0.75cm
    \caption{Returns to Education, British Estimates from the Literature.}
    \centerline{
    \begin{tabular}{c c c p{12.75cm}}
        \toprule
        OLS & IV   & Year & Reference \\ \midrule
        29.2 & 5.7 &  2010 & Dickson, M. and Smith, S., 2011. What determines the return to education: an extra year or a hurdle cleared?. Economics of education review, 30(6), pp.1167-1176. \\
        12.2 & 6.6 &  2009 & Devereux, P.J. and Fan, W., 2011. Earnings returns to the British education expansion. Economics of Education Review, 30(6), pp.1153-1166. \\
        9.6 & 5.3 &   2009 & Devereux, P.J. and Fan, W., 2011. Earnings returns to the British education expansion. Economics of Education Review, 30(6), pp.1153-1166. \\
        9.5 & 6.9 &   2006 & Grenet, J. 2013. Is it enough to increase compulsory education to raise earnings? Evidence from French and British compulsory schooling laws. Scandinavian Journal of Economics 115: 176-210. \\
        7.8 & 6.2 &   2009 & Devereux, P.J. and Fan, W., 2011. Earnings returns to the British education expansion. Economics of Education Review, 30(6), pp.1153-1166. \\
        6.0 & 4.4 &   1999 & Blanchflower, D. and Elias, P., 1999. Ability, schooling and earnings: Are twins different?. Unpublished manuscript, Warwick University. \\
        6.9 & 6.6 &   2009 & Devereux, P.J. and Fan, W., 2011. Earnings returns to the British education expansion. Economics of Education Review, 30(6), pp.1153-1166. \\
        7.0 & 7.0 &   2001 & Devereux, P.J. and R.A. Hart. 2010. Forced to be Rich? Returns to Compulsory Schooling in Britain. Economic Journal 120(549): 1345-64. \\
        7.3 & 7.7 &   1999 & Bonjour, D., Cherkas, L.F., Haskel, J.E., Hawkes, D.D. and Spector, T.D., 2003. Returns to education: Evidence from UK twins. American Economic Review, 93(5), pp.1799-1812. \\
        4.8 & 5.5 &   1991 & Dearden, Lorraine. 1999. Qualifications and earnings in Britain: how reliable are conventional OLS estimates of the returns to education? IFS Working Papers W99/07, Institute for Fiscal Studies. \\
        4.8 & 5.5 &   1981 & Dearden, L. (1998), ``Ability, Families, Education and Earnings in Britain,'' Institute for Fiscal Studies Working Paper No. W98/14. \\
        0.0 & 1.2 &   1947 & Chib, Siddhartha, and Liana Jacobi. 2016. “Bayesian Fuzzy Regression Discontinuity Analysis and Returns to Compulsory Schooling.” Journal of Applied Econometrics 31: 1026-1047. \\
        5.0 & 9.9 &   1970s & Harmon, C. and Walker, I. (2000), ``Returns to the Quantity and Quality of Education: Evidence for Men in England and Wales,'' Economica, 67: 19-35. \\
        4.6 & 10.2 &  1972 & Dickson, Matt. 2013. “The Causal Effect of Education on Wages Revisited.” Oxford Bulletin of Economics and Statistics 75 (4): 477-498. \\
        7.8 & 14.7 &  1947 & Oreopoulos, P. 2006. Estimating Average and Local Average Treatment Effects of Education when Compulsory Schooling Laws Really Matter American Economic Review96(1) 152-175 \\
        4.9 & 14.0 &  1992 & Harmon, C. and Walker, I. (1999), ``The Marginal and Average Return to Schooling in the UK,'' European Economic Review, 43(4-6), pp.879-887. \\
        6.1 & 15.3 &  1947 & Harmon, C. and I. Walker. 1995. Estimates of the economic return to schooling for the United Kingdom. American Economic Review 85: 1278-86. \\
        \bottomrule
    \end{tabular}}
    \label{tab:uk-meta}
    \justify
    \footnotesize
    \textbf{Note:}
    This table shows the education returns estimates on later-life labour market earnings in the economics literature, using an IV to causally estimate the effect of a year of education.
    Literature review kindly provided by \cite{patrinos2025causal}.
\end{table}

\end{document}